\documentclass[aps,prapplied,twocolumn,superscriptaddress]{revtex4-2}

\usepackage{amsmath,amssymb,amsfonts,amsthm}
\usepackage{graphicx}
\usepackage{hyperref}
\usepackage{enumitem}
\usepackage{bm}
\usepackage{siunitx}
\usepackage[english]{babel}

\usepackage[normalem]{ulem}

\usepackage{xcolor}

\begin{document}

\title{Engineered broadband Purcell protection using a shared $\Pi$ filter for multiplexed superconducting qubits}

\author{Samuel D. Escribano}
\affiliation{Qarakal Quantum Ltd., Derech Menachen Begin 144, Tel Aviv 6492102, Israel}

\author{Yael Kriheli}
\affiliation{Qarakal Quantum Ltd., Derech Menachen Begin 144, Tel Aviv 6492102, Israel}

\author{Samuel Goldstein}
\affiliation{Qarakal Quantum Ltd., Derech Menachen Begin 144, Tel Aviv 6492102, Israel}

\author{Daniel Dahan}
\affiliation{Qarakal Quantum Ltd., Derech Menachen Begin 144, Tel Aviv 6492102, Israel}

\author{Nadav Katz}
\email{nkatz@qarakal.ai}
\affiliation{Qarakal Quantum Ltd., Derech Menachen Begin 144, Tel Aviv 6492102, Israel}
\affiliation{Racah Institute of Physics, Hebrew University of Jerusalem, Jerusalem 9190401, Israel}


\begin{abstract}
We propose a broadband Purcell-protection scheme based on a single shared filter integrated directly into the feedline, enabling simultaneous protection of multiple qubits in a compact architecture with minimal hardware overhead. The filter consists of two open-ended stubs connected by an in-line transmission line, forming a $\Pi$ geometry, and operates via engineered passive microwave interference that suppresses the real part of the environmental admittance over a wide frequency window. Circuit simulations and finite-element modeling show that transmission is suppressed by several orders of magnitude within the target band (the qubits' frequencies) while preserving the readout mode of the multiplexed architecture. For realistic device parameters, the proposed design yields Purcell-limited relaxation times exceeding $1$~ms over a frequency span of approximately $1.5$~GHz. Our results establish the $\Pi$ filter as a compact and scalable solution for broadband impedance engineering in superconducting quantum circuits, compatible with standard dispersive readout protocols.
\end{abstract}

\maketitle

\section{Introduction} 

A Purcell filter is a circuit element designed to suppress qubit relaxation into its electromagnetic environment while preserving fast, high-fidelity readout~\cite{Reed:10, Sete:14, Krantz:19}. In circuit quantum electrodynamics (circuit-QED), this relaxation is known as Purcell decay and arises from the coupling of the qubit to dissipative modes of the readout circuitry. The associated decay rate can be expressed as $\Gamma_{\rm P} \propto \Re\{Y_{\rm env}(\omega)\}$~\cite{Houck:08,Krantz:19,Blais:21}, where $Y_{\rm env}(\omega)$ is the frequency-dependent admittance of the environment. Because the qubit and readout resonator frequencies are typically well separated, Purcell filters operate by engineering the impedance of the circuit such that the environment is strongly coupled at the readout frequency while being effectively decoupled at the qubit frequency~\cite{Sete:15}.

In multiplexed readout architectures, one strategy is to protect each readout channel with a local filtering element, either through a dedicated Purcell filter or through filtering functionality integrated into the readout circuit itself~\cite{Bronn:15,Heinsoo:18,Swiadek:24,Yen:25}. Because these filters are designed around a specific resonator or frequency channel, they can provide strong and selective suppression of radiative decay for the corresponding qubit. This local control can also be advantageous for suppressing the drive of non targeted readout resonators during multiplexed measurement, thereby reducing measurement-induced crosstalk~\cite{Heinsoo:18}. The trade-off is that the protection is tied to individual channels or resonator geometries, which increases the number of design parameters and can add footprint, routing constraints, and frequency-allocation overhead as the processor size grows.

\begin{figure}
    \centering
    \includegraphics[width=0.495\textwidth]{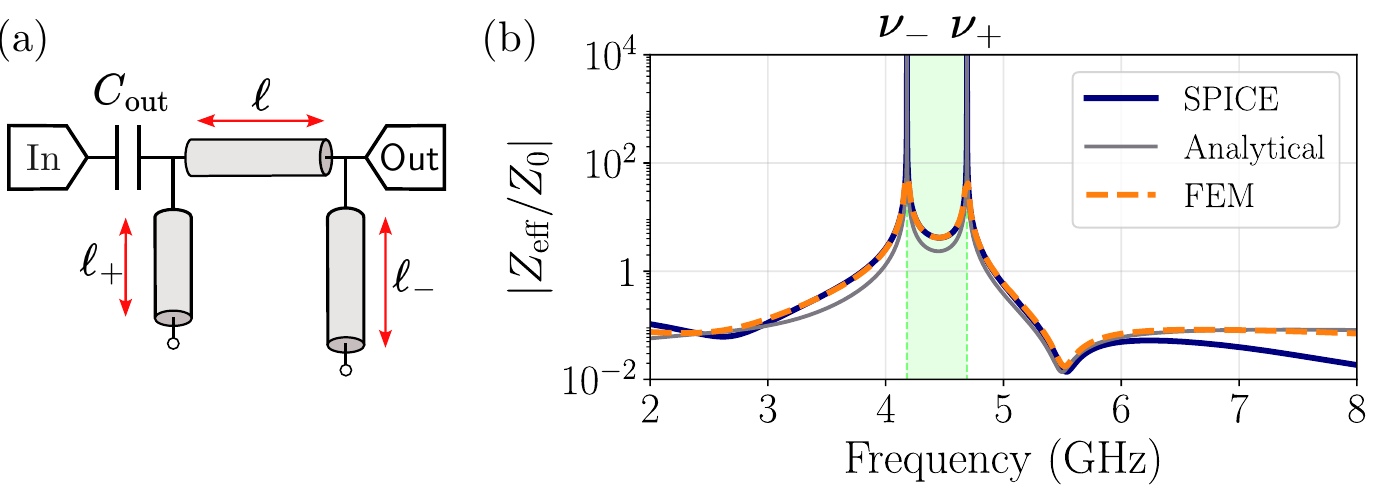}
    \caption{{\bf Isolated Purcell $\Pi$ filter.} (a) Schematic of the Purcell $\Pi$ filter, consisting of two open-ended stubs of lengths $\ell_{+}$ and $\ell_{-}$ connected by an in-line transmission line of length $\ell$ (all in gray). An output capacitor $C_{\rm out}$ connects the potential input circuit with the filter. (b) Magnitude of the effective, series impedance $|Z_{\rm eff}|$ as a function of frequency. Results obtained from SPICE simulations (dark blue), finite-element (FEM) simulations using HFSS (dashed orange), and the analytical expression of Eq.~\eqref{Eq:impedance} (gray). The analytical result applies only in the limit $C_{\rm out}\rightarrow0$. The impedance exhibits two resonances at $\nu_\pm$ associated with the electrical lengths of the stubs $\ell_\pm$. Parameters used for these simulations can be found in Appendix~\ref{Appendix:A}.}
    \label{fig:fig5a}
\end{figure}

A complementary strategy to overcome these problems is to use a shared filtering element in the measurement line. Shared bandpass Purcell filters have enabled efficient multiplexed readout while suppressing qubit decay through a common output channel~\cite{Jeffrey:14,Sete:15}. Their performance, however, is tied to the placement of a common passband and stopband relative to all readout and qubit frequencies, so the filter must be co-designed with the full frequency plan of the device. Higher-order and multistage Purcell filters improve the on--off ratio, roll-off, and usable readout bandwidth~\cite{Bronn:15,Bronn:15-2,Bronn_2017,Yan:23,Zhou:24}, but the challenge is that they introduce additional resonant sections, which increase circuit complexity, footprint, and sensitivity to parameter choices. Multimode filters can further incorporate additional functionality, such as fast reset or intrinsic Purcell protection~\cite{Xu:25}, at the cost of managing extra modes and avoiding unwanted hybridization with qubit or readout resonances. Finally, structured electromagnetic environments, such as metamaterial transmission lines~\cite{Mirhosseini:18}, provide powerful broadband impedance-engineering capabilities, although their implementation may not be compatible with every device layout or design workflow.

Motivated by these considerations, we propose a shared Purcell filter designed to balance broadband protection, low layout overhead, architectural simplicity, and compatibility with multiplexed readout. The filter is integrated into the feedline downstream of the output capacitance and consists of two open-ended stubs connected by an in-line transmission line, forming a $\Pi$ geometry [see Fig.~\ref{fig:fig5a}(a)]. This configuration extends earlier stub-based filters~\cite{Reed:10} by introducing an additional resonant element and an interconnecting transmission line. The two stub lengths set the lower and upper edges of the protected band, while the in-line section controls the interference between the corresponding scattering paths. As a result, a single passive feedline element can suppress coupling to the external environment over a finite qubit-frequency window without requiring a separate filter for each readout channel, a large number of resonant sections, or a strongly structured global transmission line.

The operation of the $\Pi$ filter can be understood as passive microwave interference. A qubit coupled to the feedline emits into a continuum of propagating modes, while the two stubs scatter this field back with frequency-dependent amplitudes and phases. At specific frequencies, set by the geometry of the filter, the emitted and scattered fields cancel exactly at the output channel, producing zeros of the radiative coupling to the external environment. When the two cancellation frequencies are placed on either side of the target qubit band, their suppression windows overlap, yielding strong Purcell protection throughout the intervening frequency range. In this way, the $\Pi$ filter provides simultaneous protection for all qubits coupled to the same feedline while leaving the readout band largely unaffected. For realistic circuit parameters, our simulations yield Purcell-limited relaxation times exceeding $1$~ms over a qubit-frequency span of approximately $1.5$~GHz.

Although the analysis presented in this work focuses on transmon qubits~\cite{Koch:07}, the $\Pi$ filter is a passive, linear microwave element and could, in principle, be incorporated into a wide range of superconducting qubit and readout architectures, including flux~\cite{Orlando:99} and fluxonium-based platforms~\cite{Manucharyan:09, Nguyen:19, Ding:23, Rower:24}, as well as dual-rail~\cite{Teoh:23, Levine:24} and cat qubits~\cite{Mirrahimi:14, Ding:25}. More generally, the same interference mechanism can be exploited to control radiative coupling in other microwave quantum devices, such as parametric amplifiers~\cite{Yamamoto:08}, quantum-limited detectors~\cite{Clerk:10}, and in general in hybrid circuit-QED systems~\cite{Kurizki:15}, where engineering the electromagnetic environment is essential for optimizing coherence and signal transfer.

\section{Purcell $\Pi$ filter: design and physical mechanism}
In this section, we present the design and operating principles of the Purcell $\Pi$ filter from two complementary perspectives. First, we analyze its response in terms of the effective impedance of the environment, showing how the filter suppresses the dissipative component of the admittance at the qubit frequency while preserving the desired readout response. We then provide a complementary interpretation based on a scattering picture of microwave interference, where the suppression of Purcell decay arises from destructive interference between the field emitted by the qubit and the fields scattered by the filter.

\subsection{Impedance engineering of the environment}
Purcell decay is determined by the real part of the admittance of the electromagnetic environment seen by the qubit. Suppressing radiative relaxation therefore requires engineering the impedance presented by the readout circuitry so that the environment becomes effectively reactive within the targeted frequency range. In the present architecture [see Fig.~\ref{fig:fig5a}(a)], the $\Pi$ filter is directly connected to the feedline and can be treated as a microwave two-port network. This representation allows us to determine the impedance presented by the filter when connected to the external environment. The response of the network is fully characterized by its ABCD transfer matrix 
\begin{equation}
    M_{\Pi} = \begin{pmatrix} A & B \\ C & D
    \end{pmatrix},
    \label{Eq:ABCD}
\end{equation}
whose matrix elements are (see Appendix~\ref{App:analytical_impedance} for a detailed derivation)
\begin{eqnarray}
A&=&\cos(\beta\omega)-\sin(\beta\omega)\tan(\beta_{-}\omega), \\
B&=&iZ_{\rm TL}\sin(\beta\omega), \\
 C&=&\frac{i}{Z_{\rm TL}}\left\{\sin(\beta\omega)\left[1-\tan(\beta_{+}\omega)\tan(\beta_{-}\omega)\right]\right. \nonumber \\
 &&\left.+\cos(\beta\omega)\left[\tan(\beta_{+}\omega)+\tan(\beta_{-}\omega)\right] \right\}, \\
 D &=& \cos(\beta\omega)-\sin(\beta\omega)\tan(\beta_{+}\omega).
\end{eqnarray}
Here $\omega$ is the angular frequency and $\beta\equiv \ell/v$ denotes the propagation delay time of the in-line transmission line while $\beta_{\pm}\equiv \ell_{\pm}/v$ corresponds to the two open-ended stubs, being $\ell$ the length of the in-line transmission line and $\ell_{\pm}$ the length of the stubs [see Fig.~\ref{fig:fig5a}(a)]. The parameter $v$ denotes the phase velocity and $Z_{\rm TL}$ the characteristic impedance of the transmission lines.

From this ABCD representation, the effective impedance presented by the $\Pi$ filter when terminated by a load impedance $Z_0$ is
\begin{widetext}
\begin{equation}
    Z_{\Pi} = Z_{\rm TL}\frac{\frac{Z_0}{Z_{\rm TL}}\left[1-\tan(\beta\omega)\tan(\beta_{-}\omega)\right]+i\tan(\beta\omega)}
    {i\frac{Z_0}{Z_{\rm TL}}
    \left\{\tan(\beta\omega)\left[1-\tan(\beta_{+}\omega)\tan(\beta_{-}\omega)\right]+\left[\tan(\beta_{+}\omega)+\tan(\beta_{-}\omega)\right] \right\} +1-\tan(\beta\omega)\tan(\beta_{+}\omega)}.
    \label{Eq:impedance}
\end{equation}
\end{widetext}
Importantly, the Purcell $\Pi$ filter does not provide protection simply by blocking current flow at selected frequencies. Instead, protection arises when the effective impedance (or equivalently the admittance) seen through the coupling network becomes purely reactive. In practice, this corresponds to frequencies at which
\begin{equation}
    \Re\left\{Z_{\Pi}\right\}\rightarrow 0,
\end{equation}
such that no real power can be dissipated into the environment. This condition is met when the denominator of $\Re\left\{Z_{\Pi}\right\}$ diverges
\begin{eqnarray}
    &&\left[1-\tan(\beta\omega)\tan(\beta_{+}\omega)\right]^2\nonumber \\
    &+&\left\{\tan(\beta\omega)\left[1-\tan(\beta_{+}\omega)\tan(\beta_{-}\omega)\right] \right. \nonumber \\
    &+&\left.\left[\tan(\beta_{+}\omega)+\tan(\beta_{-}\omega)\right] \right\}^2\rightarrow\infty.
\end{eqnarray}
This occurs at the resonant frequencies determined by the open-ended stubs, $\beta_\mp \omega_\pm = \pi/2$, which define the frequencies $\omega_\pm$ at which Purcell decay is strongly suppressed, independently of the value of $\beta$. By contrast, choosing $\beta = \beta_{+} + \beta_{-}$ produces the opposite effect, namely, $\mathrm{Re}\{Z_{\Pi}\} \rightarrow \infty$, at the mean target frequency $(\omega_++\omega_-)/2$ as a result of destructive interference between the two stubs. For this reason, we select $\beta = (\beta_{+}+\beta_{-})/2$, which ensures constructive interference of the corresponding standing-wave patterns and maximizes the suppression of Purcell decay.

To illustrate this behavior, Fig.~\ref{fig:fig5a}(b) shows the magnitude of the series impedance as a function of frequency obtained from the analytical expression (gray line), SPICE simulations (dark blue line), and finite-element (FEM) simulations (dashed orange line). For the SPICE and FEM simulations, we include a capacitance $C_{\rm out}$, as shown in Fig.~\ref{fig:fig5a}(a). All three approaches show excellent agreement, with the impedance exhibiting two divergence peaks at frequencies $\nu_\pm = \omega_\pm/2\pi$. See Appendix~\ref{Appendix:A} for further details on the simulations.

We remark that the linewidth of these peaks is governed by the ratio $Z_0/Z_{\rm TL}$. Larger values of this ratio lead to a faster divergence of the impedance near the resonance, thereby increasing the linewidth of the peaks and, in turn, the Purcell protection.

\subsection{Interference-induced Purcell suppression}

The Purcell protection provided by the $\Pi$ filter can be understood as an interference effect between the field emitted by the qubit into the feedline and the fields scattered by the filter. In the present linear regime, this interference is the classical microwave interference of fields propagating in a transmission-line network. Nevertheless, it is useful to formulate the same process in a Hamiltonian input-output language, since this places the feedline, the filter, and the qubit within a common framework that can be extended to circuits containing nonlinear quantum elements or to regimes involving strong microwave drives. For the passive linear filter considered here, this description reduces to the corresponding classical scattering response. In this picture, the qubit radiates into a continuum of propagating modes, and its decay rate is determined by the amplitude of these dressed modes at the qubit position. Suppression of the Purcell decay occurs when the emitted field is destructively interfered with the reflected fields, effectively canceling the radiative coupling to the continuum.

\begin{figure}
    \centering
    \includegraphics[width=0.33\textwidth]{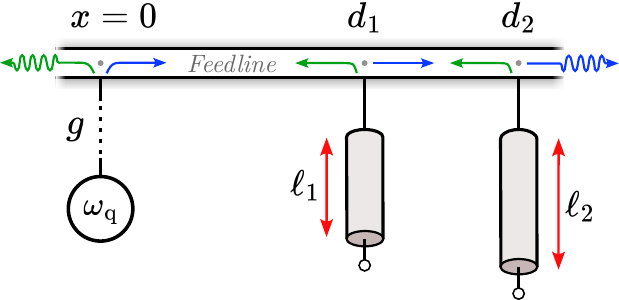}
    \caption{{\bf Microwave interference.} Sketch of the minimal model used to describe the microwave interference mechanism underlying the Purcell filter. A qubit with transition frequency $\omega_q$ is coupled to a feedline with an effective strength $g$, along which two stubs of different lengths, $\ell_1$ and $\ell_2$, are connected at positions $x = d_1$ and $x = d_2$. Under certain geometrical conditions, the signals reflected from the stubs (green arrows) interfere destructively with the outgoing field emitted by the qubit (blue arrows), thereby suppressing radiative decay into the environment (oscillating arrows), i.e., the Purcell decay. }
    \label{fig:fig6a}
\end{figure}

A minimal model capturing this mechanism consists of a qubit coupled to a one-dimensional waveguide (the feedline), with two stubs attached at positions $d_1$ and $d_2$ (see Fig.~\ref{fig:fig6a}). The total Hamiltonian reads (taking $\hbar=1$)
\begin{eqnarray}
    H &=& H_{\rm q} + H_{\rm f} + H_{\rm stubs} + H_{\rm f\text{-}stubs} + H_{\rm q\text{-}f},
\end{eqnarray}
with~\cite{Shen:05, Shen:09}
\begin{eqnarray}
    H_{\rm q} &=& \frac{\omega_q}{2}\sigma_z, \\
    H_{\rm f} &=& -i v \int dx \left[ a_R^{\dagger}(x)\partial_x a_R(x) - a_L^{\dagger}(x)\partial_x a_L(x) \right],\quad \quad \\
    H_{\rm stubs} &=& \omega_1 a_1^{\dagger} a_1 + \omega_2 a_2^{\dagger} a_2,
\end{eqnarray}
and interaction terms
\begin{eqnarray}
    H_{\rm f\text{-}stubs} &=& J_1  a_1^{\dagger}\left[a_R(d_1)+a_L(d_1)\right] \nonumber \\
    &&+ J_2 a_2^{\dagger}\left[a_R(d_2)+a_L(d_2)\right] + \mathrm{h.c.}, \\
    H_{\rm q\text{-}f} &=& g  \sigma_+ \left[a_R(0)+a_L(0)\right] + \mathrm{h.c.}
\end{eqnarray}
Here, $\omega_q$ is the qubit transition frequency and $\omega_j = \pi v/(2\ell_j)$ are the fundamental frequencies of the $j\in\left\{1,2\right\}$ stubs, set by their lengths $\ell_j$. The parameter $v$ is the propagation velocity of the transmission line. The couplings $J_{j}$ and $g$ describe the interaction of the stubs and the qubit with the waveguide, respectively. The operators $\{\sigma_{\pm}, \sigma_z\}$ are the Pauli matrices acting on the qubit subspace, while $a_{R,L}(x)$ are the bosonic annihilation operators of the right- and left-propagating modes of the waveguide and $a_{j}$ is the one of the $j$-stub mode. See Appendix~\ref{App:Quantum_interference} for an extended discussion.

This model does not explicitly include the readout resonator typically present in standard architectures, which is usually placed between the qubit and the feedline. However, the readout frequency lies well above the qubit frequency and outside the regime of interest considered here, such that the system operates in a strongly dispersive regime. As a result, the qubit and feedline are effectively coupled through an indirect interaction, which we denote by an effective coupling $g$. Including the resonator would primarily renormalize this parameter through its detuning and coupling strengths, apart from a small shift of the qubit's frequency, without qualitatively modifying the dynamics in the frequency range of interest, i.e., around the qubits's frequency. Moreover, as we show below, the $\Pi$ filter induces a frequency-selective modification of the electromagnetic environment, with negligible impact far from its resonance frequencies. In particular, the readout resonator modes remain essentially unperturbed. Therefore, neglecting the resonator in this minimal model is a reasonable approximation.

Within this model, the Purcell decay rate can be written as
\begin{equation}
    \Gamma_{\rm P} = 2\pi \rho_q\, |g|^2
    \left|
    1 +
    \left(
    r_1
    +
    \frac{t_1^2\, r_2\, e^{2ik_q(d_2-d_1)}}
    {1 - r_1 r_2 e^{2ik_q(d_2-d_1)}}
    \right)
    e^{2ik_q d_1}
    \right|^2,
\end{equation}
where $k_q = \omega_q / v$ is the momentum at the qubit frequency and $\rho_q\equiv\rho(\omega_q)$ is its density of states. This expression is written in terms of the single-stub reflection and transmission amplitudes,
\begin{eqnarray}
    r_j\equiv r_j(\omega_q) &=& -\frac{i \tan\!\left(\beta_j\omega_q\right)}{2 + i \tan\!\left(\beta_j\omega_q\right)}, \\
    t_j\equiv t_j(\omega_q) &=& \frac{2}{2 + i \tan\!\left(\beta_j\omega_q\right)},
\end{eqnarray}
with $\beta_j=\ell_j/v$ the propagation delay time of the stub $j$ and $\ell_j$ its length. These correspond to the exact transmission-line response of directly connected stubs, obtained from the boundary conditions of the circuit. Near their fundamental resonance, they reduce to the same Lorentzian form as the discrete-mode Hamiltonian, ensuring consistency between both descriptions.

In the case where only one stub is effectively coupled to the feedline ($r_2 = 0$) and its resonance is tuned to the qubit frequency, $\omega_j \simeq \omega_q$, the reflection and transmission amplitudes satisfy $r_j\simeq -1$ and $t_j\simeq 0$. In this limit, the Purcell decay rate reduces to
\begin{eqnarray}
    \Gamma_{\rm P}
    =
    2\pi \rho_q\, |g|^2
    \left| 1 - e^{2ik_q d_1} \right|^2.
\end{eqnarray}
This decay rate vanishes when $k_q d_1 = n\pi$, with $n \in \mathbb{Z}$, corresponding to complete destructive interference between the field emitted by the qubit and the field reflected by the stub.  

When both stubs contribute, broadband suppression can be achieved by placing their resonances on either side of the target qubit frequency band. Each stub suppresses the decay within a frequency window of width $\sim \kappa_j = v/\ell_j$ around its resonance. When the detuning between the two resonances is comparable to or smaller than their linewidth, $\delta\omega \lesssim \kappa/2$, these windows overlap and merge into a continuous interval. Within this interval, the destructive interference condition is not exactly satisfied at every frequency, but the decay remains strongly suppressed. As a result, the achievable bandwidth of the protection is set by the linewidth of the stub resonances, which determines the frequency range over which the interference condition can be approximately maintained.

Significantly, the mechanism described here is more general than the specific $\Pi$ filter implementation. In essence, Purcell protection can be achieved by engineering the environment such that the total reflection amplitude satisfies $-1$ over a finite frequency range, thereby maintaining destructive interference at the qubit position. The $\Pi$ filter provides a concrete realization of this idea by combining two nearby resonances whose linewidths overlap, producing near-perfect reflection across the target band. More generally, other filter designs could be devised to shape the frequency dependence of the reflected signal in a similar way, achieving broadband Purcell protection through the same interference mechanism.

\section{Purcell protection in devices}

\begin{figure}[t]
    \centering
    \includegraphics[width=0.49\textwidth]{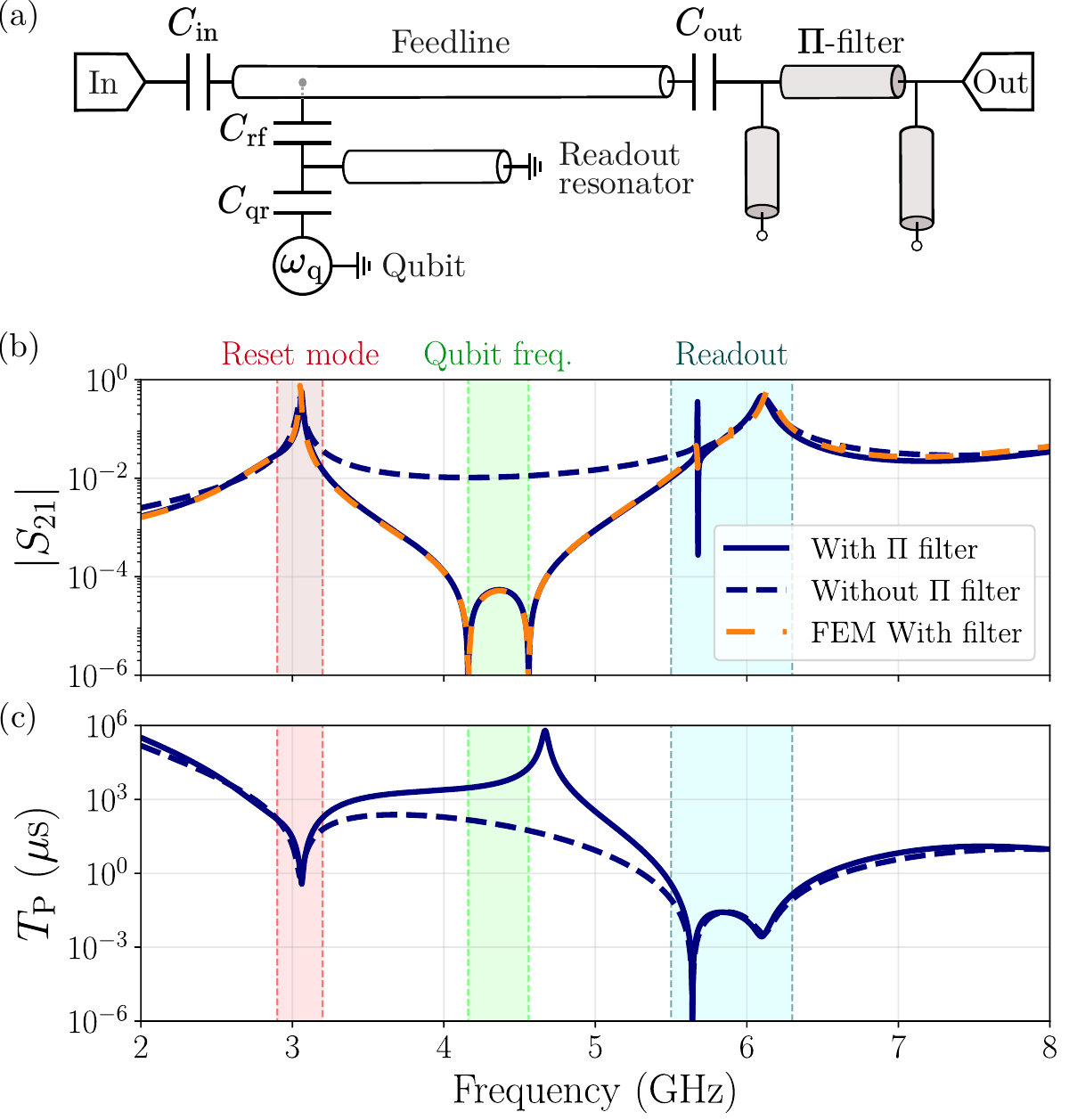}
    \caption{{\bf Purcell $\Pi$ filter protection.} (a) Circuit architecture: A qubit with a dedicated readout resonator is capacitively coupled to a transmission line (feedline). Input and output capacitors regulate the coupling to the external environment. A $\Pi$ filter (in gray) is placed after the output capacitor and consists of two open-ended stubs separated by an in-line transmission line. (b) Simulated transmission coefficient $|S_{21}|$ between the input and output ports as a function of frequency, without the $\Pi$ filter (dashed dark blue line) and with the $\Pi$ filter, using SPICE simulations (solid dark blue line) or FEM (dashed orange line). (c) Purcell relaxation time $T_{\rm P}$ of the qubit as a function of its transition frequency, with (solid line) and without (dashed line) the $\Pi$ filter. Parameters used for these simulations can be found in Appendix~\ref{Appendix:A}.}
    \label{fig:fig1}
\end{figure}

Having established the operating principle of the $\Pi$ filter, we now analyze its performance in a typical circuit-QED architecture containing a single qubit and readout resonator. The device layout is schematically illustrated in Fig.~\ref{fig:fig1}(a). The Purcell $\Pi$ filter (in gray) is integrated into the feedline immediately downstream of the output capacitance $C_{\rm out}$. Placing the $\Pi$ filter downstream of $C_{\rm out}$ separates the protected-band design from the definition of the probe-line modes: the finite probe line and its boundary capacitances set the standing-wave modes, while the external $\Pi$ filter shapes the output admittance predominantly at the target frequency band. More particularly, the lengths of the two stubs are chosen such that their fundamental modes coincide with the lowest and highest qubit frequencies along the feedline. The length of the in-line transmission line is then selected to enforce constructive interference between the responses of the two stubs; in practice, this is achieved by choosing it to be approximately the average of the stub lengths.

In Fig.~\ref{fig:fig1}(b), we present the simulated magnitude of the transmission coefficient, $|S_{21}|$, for the circuit shown in panel (a) between the input and output ports. The results obtained from numerical SPICE simulations are shown in dark blue, both with (solid line) and without (dashed line) the $\Pi$ filter. The simulations are performed in the linear-response
regime: the Josephson nonlinearity of the qubit is not included in the
scattering calculation, and the qubit/readout circuit is represented by its
linearized microwave response. Thus, the simulated $S$-parameters characterize
the passive electromagnetic environment of the device under weak probing, rather
than the nonlinear response of a strongly driven qubit. We also show the FEM simulations of the same device in orange for completeness, which exhibits an excellent agreement with the SPICE simulations.

A pronounced deviation between the two dark-blue curves appears in the frequency range 
$\left[4.2,4.5\right]$~GHz, in agreement with the target band of the filter. In this interval, 
the circuit incorporating the $\Pi$ filter exhibits a strong suppression of signal 
transmission between the input and output ports. Outside this band, the response remains 
largely unaffected. In particular, the two prominent transmission peaks around 
$6$~GHz and $3$~GHz correspond to dissipative standing-wave modes of the finite probe 
line, whose frequencies are set primarily by the probe-line length, the boundary 
capacitances, and the external impedances. We refer to the higher-frequency mode as the 
readout mode because it is designed to couple to the readout resonator and mediate 
dispersive measurement. The lower-frequency mode, in turn, has been experimentally demonstrated as a reset mode in multimode reset architectures~\cite{Xu:25}. Realizing this functionality requires appropriate engineering of the direct and indirect couplings between the qubit and the feedline, as discussed in Ref.~\cite{Xu:25}, which is beyond the scope of the present manuscript.

The preservation of the readout peak indicates that the $\Pi$ filter strongly modifies 
the probe-line response in the qubit-frequency band while only weakly perturbing the 
mode used for measurement. We have verified this quantitatively by extracting the 
readout frequency and linewidth shifts induced by the filter, finding they remain below $1~\mathrm{MHz}$ for the parameters considered 
here (not shown). Therefore, conventional dispersive readout through this mode is expected to remain 
compatible with the filter~\cite{Krantz:19,Blais:21}. More generally, the proposed 
architecture is compatible with any readout schemes that rely on spectral separation between 
the protected qubit band and the measurement band, but a systematic comparison with 
other readout protocols is beyond the scope of the current work.

In Fig.~\ref{fig:fig1}(c), we present the Purcell-limited relaxation time, estimated within the linear-response admittance approach as~\cite{Houck:08,Krantz:19,Blais:21}
\begin{equation}
    T_{\rm P} = \frac{C_{\Sigma,q}}{\Re\!\left\{Y_{\rm env}(\omega_q)\right\}},
\end{equation}
where $C_{\Sigma,q}$ is the total capacitance of the qubit and $Y_{\rm env}(\omega)$ denotes the admittance of the electromagnetic environment as seen from the qubit port. This expression applies to the capacitively coupled transmon-like circuit considered here, assuming weak coupling to a passive environment and neglecting internal loss channels. We observe an enhancement of the relaxation time by several orders of magnitude over a broad frequency window that stretches approximately from $3.5$ to $5\,\mathrm{GHz}$. In particular, within the target band between $4.2$ and $4.5\,\mathrm{GHz}$, the Purcell relaxation time exceeds $1\,\mathrm{ms}$ and reaches values above $10\,\mathrm{ms}$, indicating strong protection against radiative decay into the environment.

We note that the $|S_{21}|$ response exhibits several sharp features that are not reflected in the corresponding Purcell relaxation time. In particular, the dips in $|S_{21}|$ associated with the eigenmodes of the Purcell filter (at $4.2$ and $4.5\,\mathrm{GHz}$) correspond to a much smoother behavior in $T_{\rm P}$. This discrepancy arises because, although signal transmission from the qubit to the output port is strongly suppressed by the $\Pi$ filter, energy can still leak towards the input port. In this configuration, the Purcell relaxation rate is therefore limited by this residual decay channel, which is partially suppressed due to the asymmetry between the input and output coupling capacitances. 

\begin{figure}
    \centering
    \includegraphics[width=0.49\textwidth]{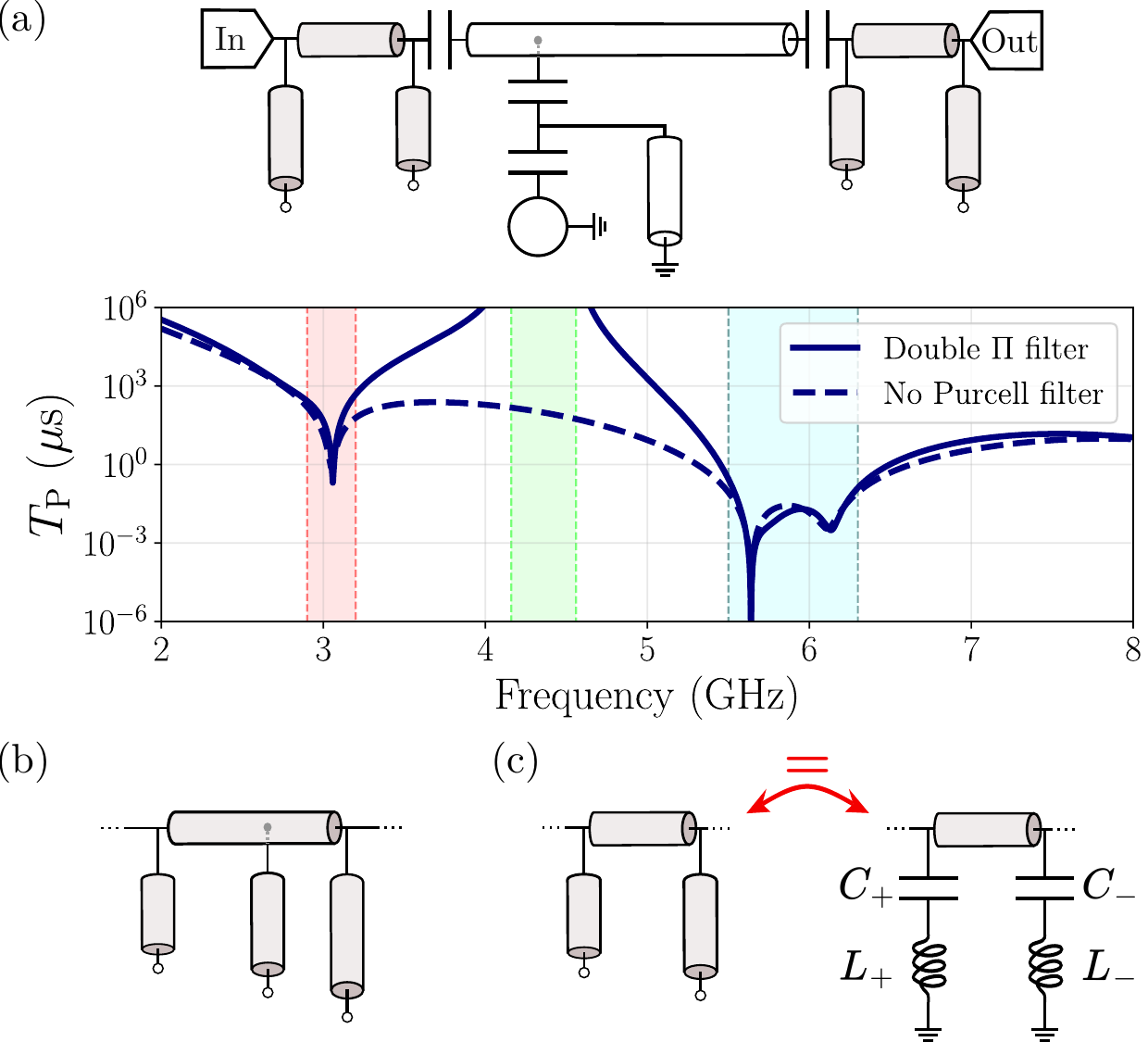}
    \caption{{\bf Alternative $\Pi$ filter architectures.} (a) Top: Schematic of a double $\Pi$ filter configuration, in which identical $\Pi$ filters are placed at both ends of the feedline; Bottom: Resulting Purcell lifetime $T_{\rm P}$, with (solid line) and without (dashed) the two filters. Parameters are the same as in Fig.~\ref{fig:fig1}. (b) Schematic of a multiplexed $\Pi$ filter composed of three (or more) stubs, enabling protection over multiple frequency bands. (c) Sketch of an equivalent $\Pi$ filter built using lumped $L$ (inductors) and $C$ (capacitors) elements.}
    \label{fig:fig5}
\end{figure}

Further suppression could be achieved by placing an identical $\Pi$ filter upstream of the input capacitor, thereby symmetrizing the electromagnetic environment and blocking radiative decay in both directions. An example of this can be found in Fig.~\ref{fig:fig5}(a). In this configuration, radiative decay channels toward both the input and output ports are suppressed. Introducing a second filter thus further reduces the real part of the environmental admittance and leads to an additional enhancement of the Purcell relaxation time (i.e., larger than $1$~s). This double-filter architecture provides a straightforward route to further improving qubit protection while preserving the simplicity of the passive interference-based design.

Another extension of the $\Pi$ filter architecture is to extend it to target multiple protected frequency windows by incorporating additional stubs, as illustrated in Fig.~\ref{fig:fig5}(b). In this multiplexed configuration, each stub is designed with a different length so that its resonance coincides with a specific qubit transition frequency or frequency band (not shown). As a result, the filter supports multiple standing-wave modes, each corresponding to a frequency at which the effective environmental admittance becomes purely reactive. This enables simultaneous Purcell protection for qubits operating in different spectral regions while preserving the global feedline architecture. Such designs may be particularly useful in large-scale processors where qubit frequencies are distributed across well-separated bands.

This generalization, however, comes at the cost of a more involved design procedure. Additional stubs introduce additional scattering paths and possible hybridization between modes, so the in-line transmission-line lengths must be chosen carefully to maintain destructive interference at the target frequencies and to avoid introducing unwanted modes in the operating bands. In Appendix~\ref{Appendix:extensions}, we study an explicit three-stub implementation and show that, with suitable numerical optimization, an additional protected band can be introduced while preserving the protection provided by the original two-stub $\Pi$ filter.

In addition, an equivalent $\Pi$ filter can be implemented using lumped $L$ and $C$ elements in place of the transmission-line stubs [see Fig.~\ref{fig:fig5}(c)]. In this configuration, the inductances $L_\pm$ and capacitances $C_\pm$ are chosen such that the resonance frequencies of the resulting $LC$ circuits match those of the transmission-line stubs with lengths $\ell_\pm$. The resulting Purcell protection is nearly identical (not shown). 
Such an implementation can reduce the physical footprint in layouts where meandered or spiral transmission-line stubs are constrained by spacing rules, parasitic coupling, or flip-chip bump grids and ultimately the minimum trace and gap dimensions imposed by the lithographic process. In addition, the lumped-element implementation is free of the undesired higher-frequency distributed modes supported by the stubs. However, lumped-element resonators introduce their own design trade-offs, including possible additional loss, parasitic inductances and capacitances, crosstalk, fabrication sensitivity, and restrictions on the achievable parameter range. The choice between distributed and lumped implementations is therefore process- and layout-dependent.

\begin{figure}
    \centering
    \includegraphics[width=0.49\textwidth]{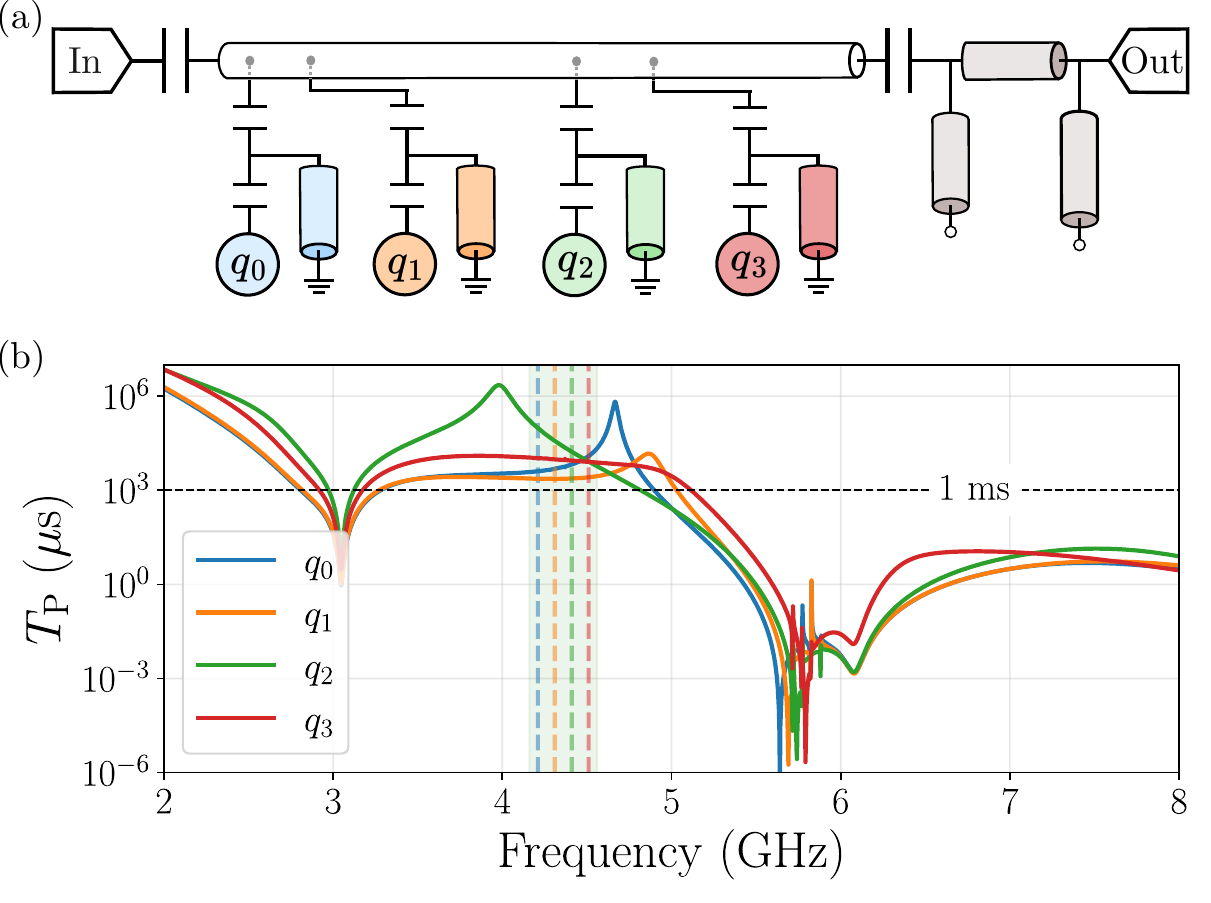}
    \caption{{\bf Purcell protection in multiplexed readout architectures.}
    (a) Multiplexed readout architecture with four qubits $q_i$, each coupled to its own readout resonator.
    (b) Purcell relaxation time $T_{\rm P}$ as a function of frequency for each qubit (different colors). The dashed vertical lines in the qubit-frequency band (shaded in green) indicate the idle frequencies of the qubits. The black dashed line marks the $1\,\mathrm{ms}$ reference level. All qubits exhibit Purcell protection exceeding $1\,\mathrm{ms}$ across a broad frequency range around their operating points. Parameters are the same as in Fig.~\ref{fig:fig1} [except for the additional qubit frequencies, which are indicated in (b)]. }
    \label{fig:fig6}
\end{figure}

Finally, to verify that the proposed $\Pi$ filter can provide simultaneous broadband protection for several qubits, we consider a multiplexed circuit-QED architecture containing several qubits coupled to a common feedline through their respective readout resonators, as illustrated in Fig.~\ref{fig:fig6}(a). In this configuration, the qubits operate at different transition frequencies (colored dashed lines) distributed within the protected band engineered by the filter (green shaded region). We simulate the complete circuit, including the feedline, multiple readout resonators, and the shared $\Pi$ filter, and compute the Purcell relaxation time for each qubit as a function of its transition frequency. The results, shown in Fig.~\ref{fig:fig6}(b), demonstrate that the filter simultaneously suppresses radiative decay for all qubits within the targeted frequency window while leaving the readout and reset modes of the multiplexed architecture largely unaffected.

To complement our study, we analyze in Appendix~\ref{Appendix:robustness} the robustness of the Purcell $\Pi$ filter against various design deviations, such as variations in the electrical lengths of the transmission lines and the output capacitance. Our results show that the Purcell protection remains unaffected under realistic design and fabrication tolerances, confirming the reliability of the proposed architecture.

\section{Conclusions}

In this work, we have introduced a Purcell filter that provides broadband protection against radiative decay. The proposed design, which we denote as a $\Pi$ filter, is placed after the output coupling capacitance in the feedline and provides simultaneous protection for all qubits coupled to it. We show analytically that the filter operates through passive interference between the field emitted by the qubit and the standing-wave modes supported by the filter, leading to destructive interference at the qubit frequencies and thereby suppressing dissipation into the external environment. At other frequencies, the filter does not significantly perturb the propagation of signals along the feedline, preserving the standing-wave structure associated with readout and control modes. As a result, the $\Pi$ filter creates broad frequency regions in which $\Re\{Y_{\mathrm{env}}(\omega)\}$ is strongly suppressed, which are tuned to fit the qubit transition frequencies, while remaining largely unaffected elsewhere.

We have performed SPICE and FEM simulations to evaluate the performance of the Purcell $\Pi$ filter in conventional circuit-QED devices. Our results demonstrate that a single $\Pi$ filter can simultaneously protect multiple qubits coupled to a common feedline in a multiplexed readout architecture. In this configuration, qubits operating at different transition frequencies within the engineered protection window exhibit strongly suppressed radiative decay, confirming that the filter provides broadband protection while preserving the integrity of the readout and reset modes. Specifically, we observe Purcell-limited relaxation times exceeding $1$~ms across the qubit frequency band, spanning more than $1$~GHz.

We have also explored extensions of the proposed architecture to address residual decay channels. While a single $\Pi$ filter suppresses radiative loss through the output port, residual leakage can still occur through the input line. This contribution can be mitigated by placing a second $\Pi$ filter before the input coupling capacitor, suppressing dissipation through both ports in turn. This symmetric configuration further isolates the qubits from the environment and enables a substantial enhancement of the Purcell-limited relaxation time, approaching the second scale. More generally, the same design principle can be extended to suppress other dissipative channels in the circuit, such as flux or voltage control lines.

Taken together, these results establish the Purcell $\Pi$ filter as a compact, scalable approach to impedance engineering in circuit quantum electrodynamics, supported by a transparent analytical model that enables systematic and predictive tailoring of its performance. Because the design relies solely on passive microwave interference and can be implemented as a single shared element 
downstream of the output capacitance, it offers a practical route to broadband Purcell protection in large-scale superconducting quantum processors. While the distributed implementation adds a non-negligible microwave length, this length is shared by all qubits coupled to the feedline. Moreover, since the filter lies outside the finite probe-line section that defines the readout modes, it can in principle be placed away from the dense qubit/readout region or implemented as part of the output microwave environment, provided that the added electrical length and parasitics are included in the design. This provides a favorable trade-off compared with dedicated Purcell filters or higher-order networks with many resonant sections.

\begin{acknowledgments}
N. K. acknowledges support from the ISF--Quantum Program.
\end{acknowledgments}

\section*{Data Availability}
The data that supports the findings of this article cannot be made publicly available because they contain commercially sensitive information. The data are available upon reasonable request from the authors.

\appendix

\section{Numerical methods and parameters}
\label{Appendix:A}
\subsection{Circuit simulations}
The circuit-level response of the Purcell $\Pi$ filter and the full circuit-QED architectures presented in this work were obtained using frequency-domain simulations based on lumped and distributed circuit models. These were performed using a SPICE-based solver (PySpice/Ngspice), and post-processing was carried out in Python.

The transmission lines composing the feedline, the $\Pi$ filter, and the readout resonators were modeled as lossless transmission line elements characterized by a phase velocity $v$ and characteristic impedance $Z_{\mathrm{TL}}$. Open-ended stubs were implemented as transmission lines terminated by open boundary conditions, thereby supporting standing-wave modes at frequencies determined by their electrical lengths.

The scattering parameters, in particular the transmission coefficient $S_{21}(\omega)$, were computed by exciting the circuit through a matched source impedance $Z_0 = 50$~$\Omega$ and solving the linear network using standard ac analysis. The effective admittance seen by the qubit, $Y_{\mathrm{env}}(\omega)$, was extracted from the input impedance at the qubit port after removing the qubit degree of freedom.

The Purcell relaxation time was then evaluated using
\begin{equation}
    T_{\rm P}(\omega_q) = \frac{C_{\Sigma,q}}{\Re\left\{Y_{\mathrm{env}}(\omega_q)\right\}},
\end{equation}
where $C_{\Sigma,\mathrm{q}}=C_{\rm q}+C_{\rm qr}$ is the total qubit's capacitance, with $C_{\rm q}$ the decoupled qubit's capacitance.

All simulations assume linear circuit elements and neglect internal losses, such that dissipation arises solely from the coupling to the external environment through the input and output ports. 
Moreover, all scattering parameters reported in this work (e.g., $S_{21}$) were computed using frequency-domain small-signal ac analysis. In this regime the circuit is linearized about its operating point, so that the transmission lines, capacitors, and resonator elements define a passive linear microwave network. The qubit nonlinearity is not included in the $S$-parameter calculation; instead, the qubit enters the Purcell analysis through the linear environmental admittance seen at the qubit port. This approximation is appropriate for extracting the passive filtering response and the associated Purcell-limited relaxation rate.

A summary of all the parameters used for our simulations can be found in Table~\ref{tab:parameters_circuit}.

\begin{table}[h]
\centering
\caption{Numerical parameters used in the circuit simulations.}
\begin{tabular}{l c c}
\hline\hline
Parameter & Symbol & Value \\
\hline
Resonator impedance & $Z_{\mathrm{RR}}$ & $69.61\,\Omega$ \\
Load impedance & $Z_0$ & $50.48\,\Omega$ \\
Transmission line permittivity & $\epsilon_{\rm eff}$ & $5.95$ \\
Qubit capacitance & $C_{\rm q}$ & $81$~fF \\
Qubit-readout capacitance & $C_{\rm qr}$ & $22$~fF \\
Readout-feedline capacitance & $C_{\rm rf}$ & $10$~fF \\
Output capacitance & $C_{\mathrm{out}}$ & $110$~fF \\
Input capacitance & $C_{\mathrm{in}}$ & $50$~fF \\
Stub lengths & $\ell_{+}$ & $6.73$~mm \\
  & $\ell_{-}$ &  $7.38$~mm \\
In-line length & $\ell$ & $7.04$~mm \\
Feedline length & $\ell_{\rm f}$ & $19.1$~mm \\
Readout frequency & $\nu_{\rm RR}$ & $5.65$~GHz \\
Qubit frequency & $\omega_{\rm q}/2\pi$ & $4.3~\mathrm{GHz}$ \\
\hline\hline
\end{tabular}
\label{tab:parameters_circuit}
\end{table}

\subsection{HFSS}

In addition to lumped element simulations, as described above, we also prepare a finite element model (FEM) to validate the SPICE results. Implemented as a computer-aided design (CAD) and exported as a graphic data system (GDS) file, this model is prepared as a fabrication-ready layout, compatible with processing in an industrial cleanroom facility, and is in fact a simplified version of an actual physical device. The layout includes several standard microwave components \cite{Krantz:19}, e.g., $50~\Omega$ launch pads for wire bonding, coplanar traveling wave guides for probing, interdigitated coupling capacitors, and rectangular spirals for the stubs realized as standing coplanar waveguides.  

Simulations are performed in Ansys Electronics Desktop (AEDT), a commercial software widely used in the superconducting qubit community for simulating  superconducting quantum devices \cite{levenson2025review, kosen2022building, savola2023design}. Only the Josephson junction itself is represented as a lumped element inductor. Notably, we assume its nonlinearity to be negligible in this context, and in any case AEDT does not support nonlinear inductances~\cite{moretti2025transmon}.  

The geometrical parameters of the interdigitated capacitors and resonators are independently calibrated using component-level simulations of the individual elements so that their extracted effective capacitances and resonance frequencies reproduce the target circuit parameters listed in Table~\ref{tab:parameters_circuit}. The resulting geometries are then used directly in the full-device simulation, without further fine-tuning of these circuit parameters. This calibration fixes the physical layout used in the HFSS simulations; the resulting full-device transmission spectra are then compared with the SPICE model without further fitting of the response curves.  The resulting transmission spectrum, displayed in Fig.~\ref{fig:fig5a}, is generated by the HFSS module, showing good agreement with the lumped element simulation.

\section{Analytical Purcell $\Pi$ filter impedance derivation}
\label{App:analytical_impedance}

In this appendix, we derive the transfer (ABCD) matrix of the Purcell $\Pi$ filter [Eq.~\eqref{Eq:ABCD}] and the associated effective impedance [Eq.~\eqref{Eq:impedance}], shown in the main text. Any reciprocal two-port network is fully characterized by its transfer (ABCD) matrix~\cite{pozar2012microwave}, which relates the voltage and current at the input port, $(V_1,I_1)$, to those at the output port, $(V_2,I_2)$,
\begin{equation}
\begin{pmatrix} V_1 \\ I_1 \end{pmatrix}
=
\begin{pmatrix} A & B \\ C & D \end{pmatrix}
\begin{pmatrix} V_2 \\ I_2 \end{pmatrix}.
\end{equation}
The main advantage of this representation is that the transfer matrix of several two-ports connected in cascade is the ordered product of the individual matrices, $M=M_1 M_2\cdots M_N$, taken from the input toward the output. When the output port is closed by a load impedance $Z_{\rm L}$, so that $V_2=Z_{\rm L}I_2$, the impedance seen from the input port is simply
\begin{equation}
Z_{\rm in}=\frac{V_1}{I_1}=\frac{A\,Z_{\rm L}+B}{C\,Z_{\rm L}+D}.
\label{Eq:Zin_ABCD}
\end{equation}

Since the $\Pi$ filter is the cascade of an open-ended shunt stub at the input node, the in-line transmission line, and a second open-ended shunt stub at the output node [see Fig.~\ref{fig:fig5a}(a)], the transfer matrix of the filter is
\begin{equation}
M_{\Pi}=M_{\rm stub}^{(+)}\,M_{\rm TL}\,M_{\rm stub}^{(-)}.
\label{Eq:Mpi_prod}
\end{equation}
A lossless transmission line of characteristic impedance $Z_{\rm TL}$, phase velocity $v$ and length $\ell$ has the transfer matrix~\cite{pozar2012microwave}
\begin{equation}
M_{\rm TL}(\theta)=
\begin{pmatrix}
\cos\left(\frac{\omega\,\ell}{v}\right) & i Z_{\rm TL}\sin\left(\frac{\omega\,\ell}{v}\right)\\[3pt]
\dfrac{i}{Z_{\rm TL}}\sin\left(\frac{\omega\,\ell}{v}\right) & \cos\left(\frac{\omega\,\ell}{v}\right)
\end{pmatrix},
\label{Eq:MTL}
\end{equation}
where $\omega$ is the angular frequency. An open-ended stub is just a transmission line terminated by an open circuit, i.e., $Z_{\rm L}\to\infty$, so its input impedance follows directly from Eqs.~\eqref{Eq:Zin_ABCD} and \eqref{Eq:MTL},
\begin{equation}
Z_{\rm stub}=\lim_{Z_{\rm L}\to\infty}\frac{A Z_{\rm L}+B}{C Z_{\rm L}+D}
=\frac{A}{C}=-\,i\,Z_{\rm TL}\cot\left(\frac{\omega\,\ell}{v}\right),
\label{Eq:Zoc}
\end{equation}
In the $\Pi$ filter the stubs are attached in shunt between the line and ground, so each one behaves as a shunt admittance $Y_{\rm stub}=1/Z_{\rm stub}$, and it is thus represented by the transfer matrix
\begin{equation}
M_{\rm stub}=
\begin{pmatrix}1 & 0\\ Y_{\rm stub} & 1\end{pmatrix},
\qquad
Y_{\rm stub}=\frac{i}{Z_{\rm TL}}\tan\left(\frac{\omega\,\ell}{v}\right).
\label{Eq:Mstub}
\end{equation}

Carrying out the matrix products with Eqs.~\eqref{Eq:MTL} and \eqref{Eq:Mstub}, and denoting $Y_{\pm}=(i/Z_{\rm TL})\tan(\beta_{\pm}\omega)$, with $\beta_j\equiv\ell_j/v$, the intermediate result is
\begin{widetext}
\begin{equation}
M_{\rm TL}(\beta\omega)\,M_{\rm stub}(\beta_{-}\omega)=
\begin{pmatrix}
\cos(\beta\omega)+iZ_{\rm TL}\sin(\beta\omega)\,Y_{-} & iZ_{\rm TL}\sin(\beta\omega)\\[3pt]
\dfrac{i}{Z_{\rm TL}}\sin(\beta\omega)+\cos(\beta\omega)\,Y_{-} & \cos(\beta\omega)
\end{pmatrix},
\end{equation}
\end{widetext}
and left-multiplying by $M_{\rm stub}(\beta_{+}\omega)$ yields the matrix elements
\begin{eqnarray}
A&=&\cos(\beta\omega)-\sin(\beta\omega)\tan(\beta_{-}\omega), \\
B&=&iZ_{\rm TL}\sin(\beta\omega), \\
C&=&\frac{i}{Z_{\rm TL}}\left\{\sin(\beta\omega)\left[1-\tan(\beta_{+}\omega)\tan(\beta_{-}\omega)\right]\right.\nonumber\\
&&\left.+\cos(\beta\omega)\left[\tan(\beta_{+}\omega)+\tan(\beta_{-}\omega)\right]\right\}, \\
D&=&\cos(\beta\omega)-\sin(\beta\omega)\tan(\beta_{+}\omega),
\end{eqnarray}
where we have used $iZ_{\rm TL}\sin(\beta\omega)Y_{\pm}=-\sin(\beta\omega)\tan(\beta_{\pm}\omega)$. These coincide with the matrix elements of Eq.~\eqref{Eq:ABCD} of the main text.

The effective impedance presented by the filter is obtained by closing its output port with the environment (load) impedance $Z_0$ and applying Eq.~\eqref{Eq:Zin_ABCD} with $Z_{\rm L}=Z_0$,
\begin{equation}
Z_{\Pi}=\frac{A\,Z_0+B}{C\,Z_0+D}.
\label{Eq:Zpi_ABCD}
\end{equation}
Inserting the matrix elements and dividing the numerator and denominator by $\cos(\beta\omega)$ gives the closed-form expression
\begin{widetext}
\begin{equation}
Z_{\Pi} = Z_{\rm TL}\frac{\frac{Z_0}{Z_{\rm TL}}\left[1-\tan(\beta\omega)\tan(\beta_{-}\omega)\right]+i\tan(\beta\omega)}
{i\frac{Z_0}{Z_{\rm TL}}
\left\{\tan(\beta\omega)\left[1-\tan(\beta_{+}\omega)\tan(\beta_{-}\omega)\right]+\left[\tan(\beta_{+}\omega)+\tan(\beta_{-}\omega)\right] \right\} +1-\tan(\beta\omega)\tan(\beta_{+}\omega)},
\end{equation}
\end{widetext}
which agrees with Eq.~\eqref{Eq:impedance} of the main text. 

We note that Eq.~\eqref{Eq:impedance} is the impedance of the $\Pi$ filter alone; in the simulated device, the filter is coupled to the feedline through the series output capacitor $C_{\rm out}$ [Fig.~\ref{fig:fig5a}(a)], which can be incorporated as an additional series block of impedance $1/(i\omega C_{\rm out})$ multiplying $M_{\Pi}$ from the input side. Nonetheless, this does not have a compact expression, and we find it more informative to show this simpler result.

\section{Scattering interpretation of Purcell suppression}
\label{App:Quantum_interference}

The Purcell protection provided by the $\Pi$ filter can also be interpreted in terms of microwave scattering in the feedline network. The interference responsible for the suppression is the classical interference of fields propagating and scattering in a linear transmission-line circuit. Nevertheless, it is useful to formulate the same process in a Hamiltonian input-output language, because this places the linear filter, the feedline, and the qubit within a common quantum-circuit framework. For the passive linear filter considered here, this description reduces to the classical scattering response derived from the transmission-line network, while providing a natural starting point for extensions involving nonlinear quantum circuit elements or strongly driven regimes.

To support this interpretation, we consider a minimal model (see Fig.~\ref{fig:fig6a} in the main text) consisting of a qubit coupled to a one-dimensional waveguide (the feedline), to which two stubs are attached at different positions. The total Hamiltonian can be written as~\cite{Shen:05, Shen:09}
\begin{eqnarray}
    H &=& H_{\rm q}+H_{\rm f}+H_{\rm stubs}+H_{\rm f-stubs} +H_{\rm q-f}, \quad\quad
\end{eqnarray}
with
\begin{eqnarray}
    H_{\rm q} &=& \frac{\omega_q}{2}\sigma_z, \\
    H_{\rm f} &=& -iv\int dx\; \left(a_R^{\dagger}(x)\partial_xa_R(x)\right. \nonumber \\
    &&\left.-a_L^{\dagger}(x)\partial_xa_L(x)\right),\\
    H_{\rm stubs} &=& \omega_1 a^{\dagger}_1a_1+ \omega_2 a^{\dagger}_2a_2, 
\end{eqnarray}
describe the qubit, the waveguide, and the two stub modes, respectively. The interaction terms are given by
\begin{eqnarray}
    H_{\rm f-stubs} &=& J_1\left[a^{\dagger}_1\left(a_R(d_1)+a_L(d_1)\right)+\mathrm{h.c.} \right] \nonumber \\
     &&+J_2\left[a^{\dagger}_2\left(a_R(d_2)+a_L(d_2)\right)+\mathrm{h.c.} \right], \quad \\
    H_{\rm q-f} &=& g\left[\sigma_+\left(a_R(0)+a_L(0)\right)+\mathrm{h.c.}\right].
\end{eqnarray}
Here, $\omega_q$ denotes the transition frequency of the qubit and $\omega_j=\pi v/2\ell_j$ is the fundamental frequencies of the two stubs (being $\ell_j$ their characteristic length). The parameter $v = 1/\sqrt{L' C'}$ is the propagation velocity in the transmission line, with $L'$ and $C'$ the inductance and capacitance per unit length. The coupling strengths $J_{1,2}$ characterize the interaction between the waveguide and each stub, while $g$ denotes the coupling between the qubit and the waveguide. In addition, to these parameters, the operators $\left\{\sigma_{\pm},\sigma_z\right\}$ act on the two-level qubit subspace, $a_{R,L}(x)$ are bosonic annihilation operators for right- and left-propagating modes of the waveguide, and $a_{1,2}$ are annihilation operators for the discrete stub modes. In this model, the qubit is taken to couple to the waveguide at $x=0$, while the two stubs are attached at positions $x=d_1$ and $x=d_2$.

The Hamiltonian above is intended as a minimal model capturing the interference mechanism underlying the Purcell protection, and therefore relies on several simplifying assumptions. We first truncate the qubit to its two lowest levels, retaining only the transition at frequency $\omega_q$ relevant for radiative relaxation. Likewise, for each stub we keep only a single resonant mode, corresponding to the frequency range of interest around the qubit transition. Higher-order stub harmonics are neglected, since they do not qualitatively modify the mechanism discussed below.

We likewise do not explicitly include the readout resonator that is typically present in standard circuit-QED architectures, where it is placed between the qubit and the feedline. This approximation is justified by the strong detuning between the readout and qubit frequencies, such that the readout mode lies outside the frequency range of interest considered here. In this regime, the qubit–feedline interaction can be effectively described by an indirect coupling, which we denote by $g$. Including the resonator would primarily result in a renormalization of this effective coupling through its detuning and coupling strengths, without qualitatively modifying the behavior in the frequency window of interest. Furthermore, as shown in the main text, the $\Pi$ filter induces a frequency-selective modification of the electromagnetic environment, with negligible impact on modes far from its resonance frequencies. In particular, the readout mode remains essentially unperturbed. Therefore, neglecting the readout resonator in this extended model constitutes a quantitatively accurate approximation.

We further assume that the waveguide is infinite, so that the electromagnetic environment is described by propagating continuum modes rather than by standing-wave modes of a finite structure. In this way, the radiative decay can be formulated naturally as a scattering problem. Within this simplified description, the detailed boundary conditions of the full feedline are not essential for understanding the origin of the protection effect, which arises from destructive interference between the field emitted by the qubit and the fields scattered by the two stubs. 

The Purcell decay rate into the bath can be computed using Fermi's golden rule, considering the one-photon transition between the excited state of the qubit and the dressed eigenstates of the environment. Since the waveguide is taken to be infinite, these states form a continuum and are intrinsically dissipative. The decay rate reads
\begin{equation}
    \Gamma_{\rm P} = 2\pi \int d\omega\; \rho(\omega)
    \left| \left\langle e,0 \middle| H \middle| g,\Psi_\omega \right\rangle \right|^2
    \delta(\omega_q - \omega),
\end{equation}
where $\rho(\omega)$ is the density of states of the waveguide, given by $\rho(\omega) = 1/(2\pi v)$, and $|\Psi_\omega\rangle$ denotes a one-photon eigenstate of the environment at frequency $\omega$.

A suitable ansatz for this state is
\begin{eqnarray}
    |\Psi_\omega\rangle &=& \int dx \; \left[
    \phi_R(x)\, a_R^\dagger(x)
    + \phi_L(x)\, a_L^\dagger(x)
    \right] |0\rangle \nonumber \\
    &&+ e_{\omega,1}\, a_1^\dagger |0\rangle
    + e_{\omega,2}\, a_2^\dagger |0\rangle,
\end{eqnarray}
where $\phi_{R,L}(x)$ are the amplitudes of the right- and left-propagating components of the wavefunction in the feedline and $e_{\omega,i}$ are the amplitudes of excitation of the $i$-th stub mode.

For the waveguide components, we take plane-wave solutions in each spatial region:
\begin{equation}
\phi_L(x)=
\begin{cases}
L_1(\omega)e^{-ikx}, & x<d_1,\\
L_2(\omega)e^{-ikx}, & d_1<x<d_2,\\
0, & x>d_2,
\end{cases}
\end{equation}
\begin{equation}
\phi_R(x)=
\begin{cases}
e^{ikx}, & x<d_1,\\
R_1(\omega)e^{ikx}, & d_1<x<d_2,\\
R_2(\omega)e^{ikx}, & x>d_2,
\end{cases}
\end{equation}
with $k=\omega/v$ the momentum of the plane-wave in the feedline, and the coefficients $L_i(\omega)$ and $R_i(\omega)$ encode all scattering processes.

These coefficients can be expressed in terms of the reflection and transmission amplitudes of each stub, $r_j(\omega)$ and $t_j(\omega)$, by summing over all possible scattering paths. For instance, the left-moving amplitude in the region $x<d_1$ is given by
\begin{eqnarray}
    L_1(\omega)
    &=& r_1(\omega)\, e^{2ikd_1} \nonumber \\
    &&+ t_1^2(\omega)\, r_2(\omega)\, e^{2ikd_2}
    \sum_{n=0}^{\infty}
    \left[r_1(\omega) r_2(\omega) e^{2ik(d_2-d_1)}\right]^n \nonumber \\
    &=& r_1(\omega)\, e^{2ikd_1}
    + \frac{t_1^2(\omega)\, r_2(\omega)\, e^{2ikd_2}}
    {1 - r_1(\omega) r_2(\omega) e^{2ik(d_2-d_1)}},
\end{eqnarray}
and similarly can be derived for the rest of the amplitudes, but they are not necessary for our derivation as shown below.

The single-stub reflection and transmission amplitudes can be obtained by solving the Schr\"odinger equation for the corresponding Hamiltonian. In the case where each stub is modeled as a single resonant mode capacitively coupled to the waveguide, they take the form
\begin{eqnarray}
    r_j(\omega) &=& -\frac{i\,\kappa_j/2}{\omega - \omega_j + i\,\kappa_j/2}, \\
    t_j(\omega) &=& \frac{\omega - \omega_j}{\omega - \omega_j + i\,\kappa_j/2},
\end{eqnarray}
with
\begin{equation}
    \kappa_j = \frac{2 J_j^2}{v},
\end{equation}
the radiative linewidth of stub $j$ into the waveguide continuum. Alternatively, if the stubs are directly connected to the transmission line (i.e., without a coupling capacitance and with a matching impedance), their scattering amplitudes can be obtained from the transmission-line boundary conditions, yielding
\begin{eqnarray}
    r_j(\omega) &=& -\frac{i \tan(\beta_j\omega)}{2 + i \tan(\beta_j\omega)}, \\
    t_j(\omega) &=& \frac{2}{2 + i \tan(\beta_j\omega)},
\end{eqnarray}
with $\beta_j = \ell_j / v$, being $\ell_j$ the length of stub $j$. In this case, the resonances lie at the same frequencies $\omega_j=\frac{\pi v}{2\ell_j}$, but the linewidth is
\begin{equation}
    \kappa_j=\frac{v}{\ell_i}.
\end{equation}

From this, the Purcell decay rate is given by
\begin{eqnarray}
    \Gamma_{\rm P} &=& 2\pi \rho(\omega_q)\, |g|^2 \, \left| \phi_{\omega_q}(0) \right|^2 \nonumber \\
    &=& 2\pi \rho(\omega_q)\, |g|^2 \, \left| 1 + L_1(\omega_q) \right|^2,
\end{eqnarray}
where $\phi_{\omega}(0)$ is the value of the dressed wavefunction at the qubit position. Using the expression for $L_1(\omega)$, this can be written explicitly as
\begin{widetext}
\begin{equation}
    \Gamma_{\rm P} = 2\pi \rho(\omega_q)\, |g|^2
    \left|
    1 +
    \left(
    r_1(\omega_q)
    +
    \frac{t_1^2(\omega_q)\, r_2(\omega_q)\, e^{2ik_q(d_2-d_1)}}
    {1 - r_1(\omega_q) r_2(\omega_q) e^{2ik_q(d_2-d_1)}}
    \right)
    e^{2ik_q d_1}
    \right|^2,
\end{equation}
\end{widetext}
with $k_q = \omega_q / v$.

In the case where one (and only one) of the stubs is tuned to the qubit frequency, $\omega_j \simeq \omega_q$, the corresponding reflection and transmission amplitudes satisfy $r_j(\omega_q) \simeq -1$ and $t_j(\omega_q) \simeq 0$. In this limit, the expression for the Purcell decay rate reduces to
\begin{eqnarray}
    \Gamma_{\rm P}
    =
    2\pi \rho(\omega_q)\, |g|^2
    \left| 1 - e^{2ik_q d_1} \right|^2.
\end{eqnarray}
This decay rate vanishes exactly when $k_q d_1 = n\pi$, with $n \in \mathbb{Z}$, corresponding to complete destructive interference between the directly emitted field and the field reflected by the stub.

Away from these exact cancellation points, the suppression of the Purcell decay is controlled by the linewidth of the stub resonance. Expanding the reflection amplitude around $\omega_j$, one finds that the condition for destructive interference is satisfied only within a frequency window of width approximately $\kappa_j$, set by the radiative linewidth of the stub. As a result, a single stub provides Purcell protection only over a narrow spectral region around its resonance frequency and cannot achieve broadband suppression.

Broadband suppression can instead be achieved by engineering the total amplitude $1 + L_1(\omega)$ of the structure such that its modulus remains small over a finite frequency interval. Equivalently, this corresponds to designing the Purcell filter so that the total reflection amplitude satisfies $L_1(\omega) \simeq -1$ across a continuous range of frequencies (the qubit frequency band), leading to destructive interference at the qubit position.

In the case of our $\Pi$ filter design, we introduce two stubs with resonance frequencies $\omega_{1,2} = \omega_0 \pm \delta\omega$, centered around the target qubit frequency $\omega_0$ (the mean qubit frequency along the feedline), and choose their detuning $\delta\omega$ to be of the order of their linewidths, $\delta\omega \sim \kappa/2$. In this regime, the corresponding reflection features overlap in frequency, and the individual suppression windows associated with each stub merge into a continuous interval. Within this interval, the interference condition
\begin{equation}
    1 + L_1(\omega) \simeq 0,
\end{equation}
is approximately maintained, resulting in a broadband suppression of the Purcell decay.

In our device, the linewidth of the stub resonances is larger than the frequency range spanned by the qubits. As a consequence, the resulting protection window exceeds the qubit band, ensuring simultaneous Purcell suppression for all qubits within the operating range.

\section{Robustness of the Purcell $\Pi$ filter}
\label{Appendix:robustness}

\begin{figure}[t]
    \centering
    \includegraphics[width=0.49\textwidth]{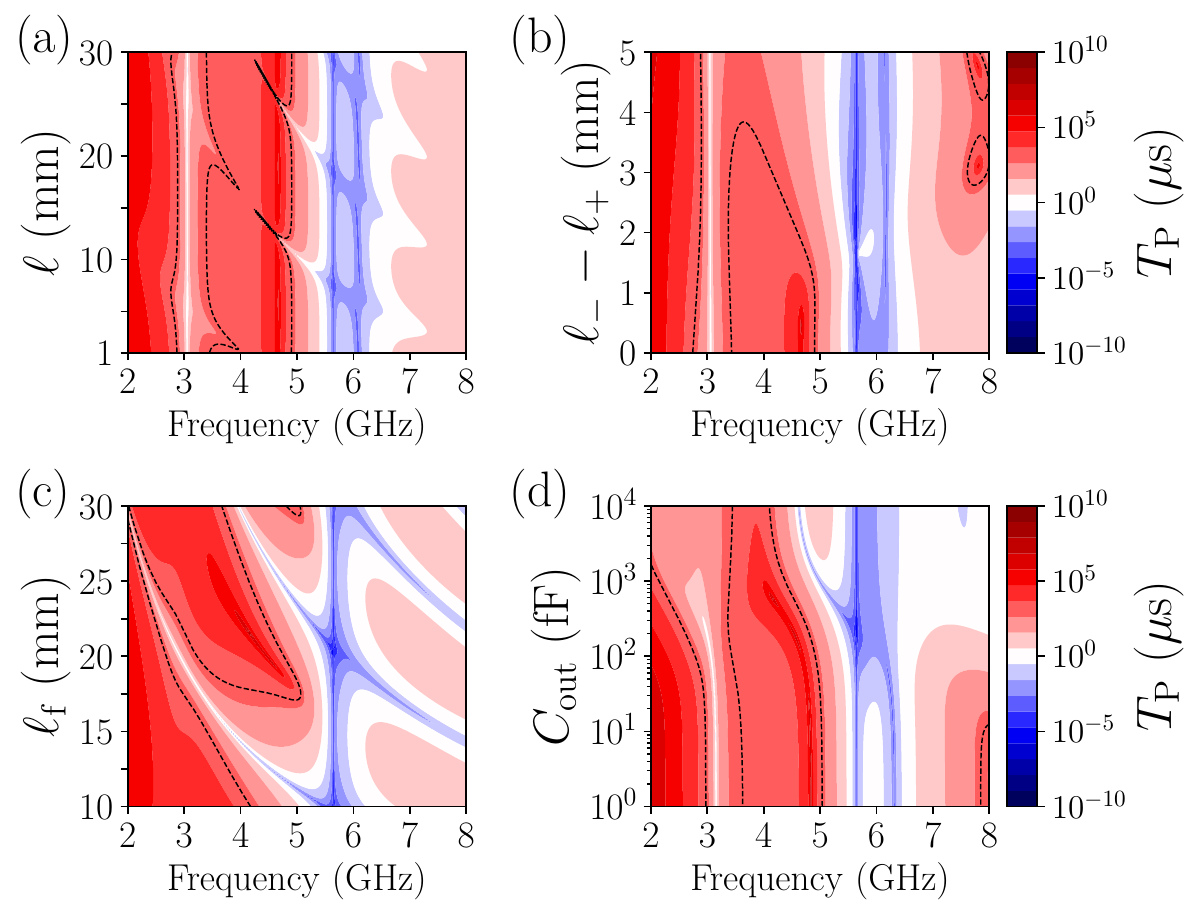}
    \caption{{\bf Design robustness.} Purcell decay as a function of frequency (horizontal axis) and selected circuit parameters (vertical axis): (a) the separation between the stubs of the $\Pi$ filter, $\ell$; (b) the difference in stub lengths, $ \ell_{\rm -}-\ell_{\rm +}$; (c) the feedline length, $\ell_{\rm f}$; and (d) the output coupling capacitance, $C_{\rm out}$. The dashed contour indicates the $1$~ms Purcell lifetime threshold. } 
    \label{fig:fig3}
\end{figure}

To assess the robustness of the proposed Purcell $\Pi$ filter against parameter variations and fabrication imperfections, we systematically analyze the dependence of the Purcell relaxation time $T_{\rm P}$ on key geometrical and circuit parameters. The results are summarized in Fig.~\ref{fig:fig3}, where each panel corresponds to a sweep of a single parameter while keeping all others fixed.

Figure~\ref{fig:fig3}(a) shows the Purcell relaxation time as a function of frequency and in-line transmission line length $\ell$ of the Purcell filter. The optimal condition $\ell \approx (\ell_+ + \ell_-)/2$ corresponds to constructive interference between the standing-wave modes of the two stubs and seems to hold for a dozen mm.

In Fig.~\ref{fig:fig3}(b), we study the effect of varying the difference between the stub lengths, $\delta \ell \equiv \ell_- - \ell_+$, which determines the width of the protected frequency window. As expected, increasing $\delta \ell$ broadens the separation between the two characteristic frequencies $\omega_\pm$, thereby widening the frequency range over which the environmental admittance becomes purely reactive. Conversely, reducing $\delta \ell$ narrows the protected band. Significantly, the results show that the filter maintains a well-defined suppression region even for moderate asymmetries, confirming that the design is robust against fabrication-induced variations in stub lengths.

Figure~\ref{fig:fig3}(c) shows the dependence of $T_{\rm P}$ on the feedline length. This parameter primarily sets the global mode structure of the circuit, including the positions of the readout and reset resonances. The Purcell filter provides effective protection only when its eigenmodes do not coincide with those of the feedline, which appear as dispersive features in panel (c).

Finally, Fig.~\ref{fig:fig3}(d) shows the dependence of the Purcell relaxation time on the output coupling capacitance $C_{\mathrm{out}}$, which controls the coupling to the external environment. Increasing $C_{\mathrm{out}}$ enhances the overall coupling strength and therefore increases the baseline dissipation. However, within the protected frequency window, the $\Pi$ filter continues to suppress the real part of the environmental admittance, preserving long relaxation times over several orders of magnitude variation in $C_{\mathrm{out}}$.

\section{Extension to multi stub Purcell filters}
\label{Appendix:extensions}

\begin{figure}[t]
    \centering
    \includegraphics[width=0.49\textwidth]{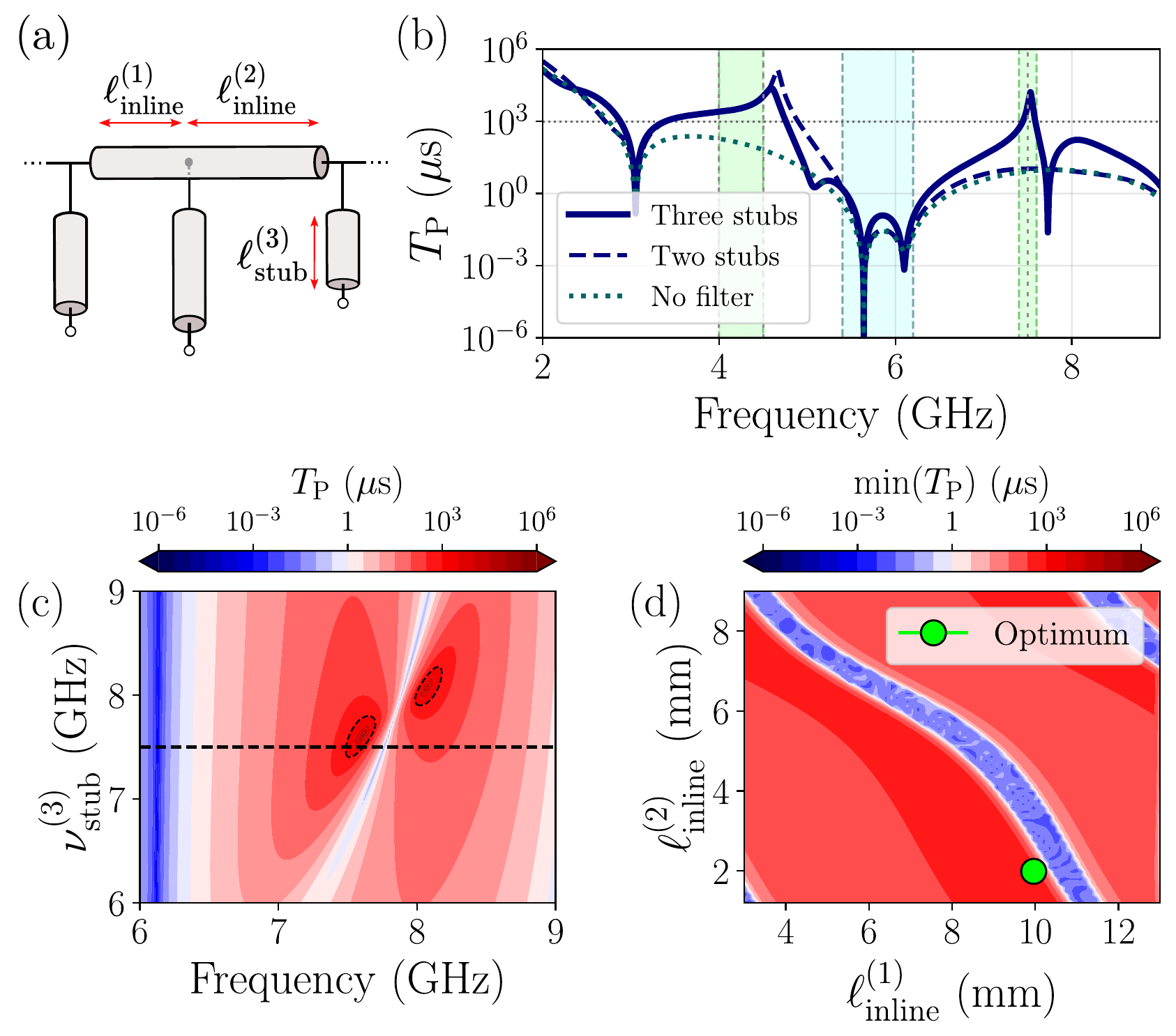}
    \caption{{\bf Three-stub Purcell $\Pi$ filter.} (a) Schematic of an extended Purcell $\Pi$ filter including a third branch. The additional branch consists of an open-ended stub of length $\ell_{\rm stub}^{(3)}$, connected to the original two-stub structure through an in-line transmission-line section of length $\ell_{\rm inline}^{(2)}$.
(b) Purcell-limited relaxation time $T_{\rm P}$ as a function of frequency without a Purcell filter (dotted line), with the two-stub Purcell $\Pi$ filter studied in the main text (dashed line), and with the three-stub filter shown in panel (a) (solid line). The target frequency regions are indicated by the green shaded areas, while the readout-frequency band is indicated by the blue shaded area.
(c) Purcell-limited relaxation time as a function of the additional-stub resonance frequency $\nu_{\rm stub}^{(3)}=c/4\ell_{\rm stub}^{(3)}\sqrt{\epsilon_{\rm eff}}$ and the qubit frequency. The dashed horizontal line indicates the value $\nu_{\rm stub}^{(3)}=7.5~\mathrm{GHz}$ used in the simulations of panels (b) and (d).
(d) Purcell-limited relaxation time as a function of the two in-line transmission-line lengths. Since the goal is to optimize protection at all targeted stub frequencies, we plot the minimum value of $T_{\rm P}$ between those target frequencies ($\nu_q$ corresponds to the qubit frequency band lying between $4$ and $4.5$~GHz that is targeted with the other two stubs). The optimal design corresponds to the maximum of this quantity, marked with a green dot.} 
    \label{fig:figAppC}
\end{figure}

The Purcell $\Pi$ filter discussed in the main text can be extended by adding additional open-ended stubs to target more than two frequencies, as we illustrated in Fig.~\ref{fig:fig5}(b). This provides a straightforward way to broaden the protected region or to protect several separated qubit-frequency bands while preserving the same basic design principle. In this appendix, we illustrate this idea with a three-stub implementation, shown schematically in Fig.~\ref{fig:figAppC}(a).

The additional branch consists of an open-ended stub with length $\ell_{\rm stub}^{(3)}$, connected to the original two-stub structure through a second in-line transmission-line section of length $\ell_{\rm inline}^{(2)}$. The resonance frequency of the third stub defines an additional frequency at which the radiative coupling to the output environment is suppressed. The in-line sections then determine how the scattering responses of the different stubs interfere with one another. As in the two-stub case, the design objective is not only to create isolated suppression points but also to obtain a broad frequency region in which the Purcell-limited lifetime remains large.

In principle, the additional branch can be used to broaden and strengthen the protection around the original qubit-frequency band. In the parameter regime considered in the main text, however, this region already exhibits large Purcell-limited lifetimes, exceeding the millisecond scale over a bandwidth larger than $1~\mathrm{GHz}$. We therefore focus instead on using the third branch to protect an additional, spectrally separated frequency band, which could be relevant for qubits, couplers, or other circuit modes operating outside the main protected window.

Figure~\ref{fig:figAppC}(b) shows the Purcell-limited relaxation time $T_{\rm P}$ as a function of frequency for a case in which the additional stub targets $\nu_{\rm stub}^{(3)}=7.5~\mathrm{GHz}$. We compare three configurations: the circuit without a Purcell filter (dotted line), the two-stub $\Pi$ filter studied in the main text (dashed line), and the three-stub extension shown in Fig.~\ref{fig:figAppC}(a) (solid line). The three-stub filter introduces an additional protected region associated with the resonance of the added stub while preserving the broadband protection produced by the original two-stub structure. In particular, around $\nu_{\rm stub}^{(3)}$, the three-stub device increases $T_{\rm P}$ by approximately three orders of magnitude relative to the unfiltered case and exceeds $1~\mathrm{ms}$ over a bandwidth of about $250~\mathrm{MHz}$.

This improvement comes at the cost of a more involved design procedure. In the two-stub $\Pi$ filter, the stub lengths directly set the target suppression frequencies, while the in-line length controls the interference between the two scattering paths. In the three-stub extension, there are now two in-line transmission-line lengths, and their values must be chosen so that the protection is simultaneously optimized at all targeted frequencies. In practice, this requires a numerical optimization, since no simple compact analytical condition is available for the multi stub case. Figures~\ref{fig:figAppC}(c) and ~\ref{fig:figAppC}(d) illustrate this design procedure.

The first step is to choose the resonance frequency of the additional stub. Figure~\ref{fig:figAppC}(c) shows $T_{\rm P}$ as a function of the third-stub resonance frequency $\nu_{\rm stub}^{(3)}=c/4\ell_{\rm stub}^{(3)}\sqrt{\epsilon_{\rm eff}}$ and the sampled frequency. As expected, the region of enhanced Purcell protection follows the resonance of the added stub. For the simulations shown in panels (b) and (d), we choose $\nu_{\rm stub}^{(3)}=7.5~\mathrm{GHz}$, indicated by the dashed line.

Once this target frequency is fixed, the in-line lengths are optimized. Figure~\ref{fig:figAppC}(d) shows the dependence of the protection on the two in-line transmission-line lengths. Since the goal is to protect all targeted frequencies simultaneously, we plot the minimum value of $T_{\rm P}$ evaluated at the stub-target frequencies. The optimal design corresponds to the maximum of this quantity, ensuring that no target frequency is protected at the expense of another. This optimization illustrates that, although additional stubs increase the number of design parameters, the design remains governed by simple and interpretable rules: the stub lengths set the target suppression frequencies, while the in-line lengths control the interference between the corresponding scattering paths.


\begin{thebibliography}{37}%
\makeatletter
\providecommand \@ifxundefined [1]{%
 \@ifx{#1\undefined}
}%
\providecommand \@ifnum [1]{%
 \ifnum #1\expandafter \@firstoftwo
 \else \expandafter \@secondoftwo
 \fi
}%
\providecommand \@ifx [1]{%
 \ifx #1\expandafter \@firstoftwo
 \else \expandafter \@secondoftwo
 \fi
}%
\providecommand \natexlab [1]{#1}%
\providecommand \enquote  [1]{``#1''}%
\providecommand \bibnamefont  [1]{#1}%
\providecommand \bibfnamefont [1]{#1}%
\providecommand \citenamefont [1]{#1}%
\providecommand \href@noop [0]{\@secondoftwo}%
\providecommand \href [0]{\begingroup \@sanitize@url \@href}%
\providecommand \@href[1]{\@@startlink{#1}\@@href}%
\providecommand \@@href[1]{\endgroup#1\@@endlink}%
\providecommand \@sanitize@url [0]{\catcode `\\12\catcode `\$12\catcode `\&12\catcode `\#12\catcode `\^12\catcode `\_12\catcode `\%12\relax}%
\providecommand \@@startlink[1]{}%
\providecommand \@@endlink[0]{}%
\providecommand \url  [0]{\begingroup\@sanitize@url \@url }%
\providecommand \@url [1]{\endgroup\@href {#1}{\urlprefix }}%
\providecommand \urlprefix  [0]{URL }%
\providecommand \Eprint [0]{\href }%
\providecommand \doibase [0]{https://doi.org/}%
\providecommand \selectlanguage [0]{\@gobble}%
\providecommand \bibinfo  [0]{\@secondoftwo}%
\providecommand \bibfield  [0]{\@secondoftwo}%
\providecommand \translation [1]{[#1]}%
\providecommand \BibitemOpen [0]{}%
\providecommand \bibitemStop [0]{}%
\providecommand \bibitemNoStop [0]{.\EOS\space}%
\providecommand \EOS [0]{\spacefactor3000\relax}%
\providecommand \BibitemShut  [1]{\csname bibitem#1\endcsname}%
\let\auto@bib@innerbib\@empty
\bibitem [{\citenamefont {Reed}\ \emph {et~al.}(2010)\citenamefont {Reed}, \citenamefont {Johnson}, \citenamefont {Houck}, \citenamefont {DiCarlo}, \citenamefont {Chow}, \citenamefont {Schuster}, \citenamefont {Frunzio},\ and\ \citenamefont {Schoelkopf}}]{Reed:10}%
  \BibitemOpen
  \bibfield  {author} {\bibinfo {author} {\bibfnamefont {M.~D.}\ \bibnamefont {Reed}}, \bibinfo {author} {\bibfnamefont {B.~R.}\ \bibnamefont {Johnson}}, \bibinfo {author} {\bibfnamefont {A.~A.}\ \bibnamefont {Houck}}, \bibinfo {author} {\bibfnamefont {L.}~\bibnamefont {DiCarlo}}, \bibinfo {author} {\bibfnamefont {J.~M.}\ \bibnamefont {Chow}}, \bibinfo {author} {\bibfnamefont {D.~I.}\ \bibnamefont {Schuster}}, \bibinfo {author} {\bibfnamefont {L.}~\bibnamefont {Frunzio}},\ and\ \bibinfo {author} {\bibfnamefont {R.~J.}\ \bibnamefont {Schoelkopf}},\ }\bibfield  {title} {\bibinfo {title} {Fast reset and suppressing spontaneous emission of a superconducting qubit},\ }\href {https://doi.org/10.1063/1.3435463} {\bibfield  {journal} {\bibinfo  {journal} {Applied Physics Letters}\ }\textbf {\bibinfo {volume} {96}},\ \bibinfo {pages} {203110} (\bibinfo {year} {2010})}\BibitemShut {NoStop}%
\bibitem [{\citenamefont {Sete}\ \emph {et~al.}(2014)\citenamefont {Sete}, \citenamefont {Gambetta},\ and\ \citenamefont {Korotkov}}]{Sete:14}%
  \BibitemOpen
  \bibfield  {author} {\bibinfo {author} {\bibfnamefont {E.~A.}\ \bibnamefont {Sete}}, \bibinfo {author} {\bibfnamefont {J.~M.}\ \bibnamefont {Gambetta}},\ and\ \bibinfo {author} {\bibfnamefont {A.~N.}\ \bibnamefont {Korotkov}},\ }\bibfield  {title} {\bibinfo {title} {Purcell effect with microwave drive: Suppression of qubit relaxation rate},\ }\href {https://doi.org/10.1103/PhysRevB.89.104516} {\bibfield  {journal} {\bibinfo  {journal} {Phys. Rev. B}\ }\textbf {\bibinfo {volume} {89}},\ \bibinfo {pages} {104516} (\bibinfo {year} {2014})}\BibitemShut {NoStop}%
\bibitem [{\citenamefont {Krantz}\ \emph {et~al.}(2019)\citenamefont {Krantz}, \citenamefont {Kjaergaard}, \citenamefont {Yan}, \citenamefont {Orlando}, \citenamefont {Gustavsson},\ and\ \citenamefont {Oliver}}]{Krantz:19}%
  \BibitemOpen
  \bibfield  {author} {\bibinfo {author} {\bibfnamefont {P.}~\bibnamefont {Krantz}}, \bibinfo {author} {\bibfnamefont {M.}~\bibnamefont {Kjaergaard}}, \bibinfo {author} {\bibfnamefont {F.}~\bibnamefont {Yan}}, \bibinfo {author} {\bibfnamefont {T.~P.}\ \bibnamefont {Orlando}}, \bibinfo {author} {\bibfnamefont {S.}~\bibnamefont {Gustavsson}},\ and\ \bibinfo {author} {\bibfnamefont {W.~D.}\ \bibnamefont {Oliver}},\ }\bibfield  {title} {\bibinfo {title} {A quantum engineer's guide to superconducting qubits},\ }\href {https://doi.org/10.1063/1.5089550} {\bibfield  {journal} {\bibinfo  {journal} {Applied Physics Reviews}\ }\textbf {\bibinfo {volume} {6}},\ \bibinfo {pages} {021318} (\bibinfo {year} {2019})}\BibitemShut {NoStop}%
\bibitem [{\citenamefont {Houck}\ \emph {et~al.}(2008)\citenamefont {Houck}, \citenamefont {Schreier}, \citenamefont {Johnson}, \citenamefont {Chow}, \citenamefont {Koch}, \citenamefont {Gambetta}, \citenamefont {Schuster}, \citenamefont {Frunzio}, \citenamefont {Devoret}, \citenamefont {Girvin},\ and\ \citenamefont {Schoelkopf}}]{Houck:08}%
  \BibitemOpen
  \bibfield  {author} {\bibinfo {author} {\bibfnamefont {A.~A.}\ \bibnamefont {Houck}}, \bibinfo {author} {\bibfnamefont {J.~A.}\ \bibnamefont {Schreier}}, \bibinfo {author} {\bibfnamefont {B.~R.}\ \bibnamefont {Johnson}}, \bibinfo {author} {\bibfnamefont {J.~M.}\ \bibnamefont {Chow}}, \bibinfo {author} {\bibfnamefont {J.}~\bibnamefont {Koch}}, \bibinfo {author} {\bibfnamefont {J.~M.}\ \bibnamefont {Gambetta}}, \bibinfo {author} {\bibfnamefont {D.~I.}\ \bibnamefont {Schuster}}, \bibinfo {author} {\bibfnamefont {L.}~\bibnamefont {Frunzio}}, \bibinfo {author} {\bibfnamefont {M.~H.}\ \bibnamefont {Devoret}}, \bibinfo {author} {\bibfnamefont {S.~M.}\ \bibnamefont {Girvin}},\ and\ \bibinfo {author} {\bibfnamefont {R.~J.}\ \bibnamefont {Schoelkopf}},\ }\bibfield  {title} {\bibinfo {title} {Controlling the spontaneous emission of a superconducting transmon qubit},\ }\href {https://doi.org/10.1103/PhysRevLett.101.080502} {\bibfield  {journal} {\bibinfo  {journal} {Phys. Rev. Lett.}\ }\textbf {\bibinfo {volume} {101}},\ \bibinfo {pages} {080502} (\bibinfo {year} {2008})}\BibitemShut {NoStop}%
\bibitem [{\citenamefont {Blais}\ \emph {et~al.}(2021)\citenamefont {Blais}, \citenamefont {Grimsmo}, \citenamefont {Girvin},\ and\ \citenamefont {Wallraff}}]{Blais:21}%
  \BibitemOpen
  \bibfield  {author} {\bibinfo {author} {\bibfnamefont {A.}~\bibnamefont {Blais}}, \bibinfo {author} {\bibfnamefont {A.~L.}\ \bibnamefont {Grimsmo}}, \bibinfo {author} {\bibfnamefont {S.~M.}\ \bibnamefont {Girvin}},\ and\ \bibinfo {author} {\bibfnamefont {A.}~\bibnamefont {Wallraff}},\ }\bibfield  {title} {\bibinfo {title} {Circuit quantum electrodynamics},\ }\href {https://doi.org/10.1103/RevModPhys.93.025005} {\bibfield  {journal} {\bibinfo  {journal} {Rev. Mod. Phys.}\ }\textbf {\bibinfo {volume} {93}},\ \bibinfo {pages} {025005} (\bibinfo {year} {2021})}\BibitemShut {NoStop}%
\bibitem [{\citenamefont {Sete}\ \emph {et~al.}(2015)\citenamefont {Sete}, \citenamefont {Martinis},\ and\ \citenamefont {Korotkov}}]{Sete:15}%
  \BibitemOpen
  \bibfield  {author} {\bibinfo {author} {\bibfnamefont {E.~A.}\ \bibnamefont {Sete}}, \bibinfo {author} {\bibfnamefont {J.~M.}\ \bibnamefont {Martinis}},\ and\ \bibinfo {author} {\bibfnamefont {A.~N.}\ \bibnamefont {Korotkov}},\ }\bibfield  {title} {\bibinfo {title} {Quantum theory of a bandpass purcell filter for qubit readout},\ }\href {https://doi.org/10.1103/PhysRevA.92.012325} {\bibfield  {journal} {\bibinfo  {journal} {Phys. Rev. A}\ }\textbf {\bibinfo {volume} {92}},\ \bibinfo {pages} {012325} (\bibinfo {year} {2015})}\BibitemShut {NoStop}%
\bibitem [{\citenamefont {Bronn}\ \emph {et~al.}(2015{\natexlab{a}})\citenamefont {Bronn}, \citenamefont {Magesan}, \citenamefont {Masluk}, \citenamefont {Chow}, \citenamefont {Gambetta},\ and\ \citenamefont {Steffen}}]{Bronn:15}%
  \BibitemOpen
  \bibfield  {author} {\bibinfo {author} {\bibfnamefont {N.~T.}\ \bibnamefont {Bronn}}, \bibinfo {author} {\bibfnamefont {E.}~\bibnamefont {Magesan}}, \bibinfo {author} {\bibfnamefont {N.~A.}\ \bibnamefont {Masluk}}, \bibinfo {author} {\bibfnamefont {J.~M.}\ \bibnamefont {Chow}}, \bibinfo {author} {\bibfnamefont {J.~M.}\ \bibnamefont {Gambetta}},\ and\ \bibinfo {author} {\bibfnamefont {M.}~\bibnamefont {Steffen}},\ }\bibfield  {title} {\bibinfo {title} {Reducing spontaneous emission in circuit quantum electrodynamics by a combined readout/filter technique},\ }\href {https://doi.org/10.1109/TASC.2015.2456109} {\bibfield  {journal} {\bibinfo  {journal} {IEEE Transactions on Applied Superconductivity}\ }\textbf {\bibinfo {volume} {25}},\ \bibinfo {pages} {1} (\bibinfo {year} {2015}{\natexlab{a}})}\BibitemShut {NoStop}%
\bibitem [{\citenamefont {Heinsoo}\ \emph {et~al.}(2018)\citenamefont {Heinsoo}, \citenamefont {Andersen}, \citenamefont {Remm}, \citenamefont {Krinner}, \citenamefont {Walter}, \citenamefont {Salath\'e}, \citenamefont {Gasparinetti}, \citenamefont {Besse}, \citenamefont {Poto\ifmmode~\check{c}\else \v{c}\fi{}nik}, \citenamefont {Wallraff},\ and\ \citenamefont {Eichler}}]{Heinsoo:18}%
  \BibitemOpen
  \bibfield  {author} {\bibinfo {author} {\bibfnamefont {J.}~\bibnamefont {Heinsoo}}, \bibinfo {author} {\bibfnamefont {C.~K.}\ \bibnamefont {Andersen}}, \bibinfo {author} {\bibfnamefont {A.}~\bibnamefont {Remm}}, \bibinfo {author} {\bibfnamefont {S.}~\bibnamefont {Krinner}}, \bibinfo {author} {\bibfnamefont {T.}~\bibnamefont {Walter}}, \bibinfo {author} {\bibfnamefont {Y.}~\bibnamefont {Salath\'e}}, \bibinfo {author} {\bibfnamefont {S.}~\bibnamefont {Gasparinetti}}, \bibinfo {author} {\bibfnamefont {J.-C.}\ \bibnamefont {Besse}}, \bibinfo {author} {\bibfnamefont {A.}~\bibnamefont {Poto\ifmmode~\check{c}\else \v{c}\fi{}nik}}, \bibinfo {author} {\bibfnamefont {A.}~\bibnamefont {Wallraff}},\ and\ \bibinfo {author} {\bibfnamefont {C.}~\bibnamefont {Eichler}},\ }\bibfield  {title} {\bibinfo {title} {Rapid high-fidelity multiplexed readout of superconducting qubits},\ }\href {https://doi.org/10.1103/PhysRevApplied.10.034040} {\bibfield  {journal} {\bibinfo  {journal} {Phys. Rev. Appl.}\ }\textbf {\bibinfo {volume} {10}},\ \bibinfo {pages} {034040} (\bibinfo {year} {2018})}\BibitemShut {NoStop}%
\bibitem [{\citenamefont {Swiadek}\ \emph {et~al.}(2024)\citenamefont {Swiadek}, \citenamefont {Shillito}, \citenamefont {Magnard}, \citenamefont {Remm}, \citenamefont {Hellings}, \citenamefont {Lacroix}, \citenamefont {Ficheux}, \citenamefont {Zanuz}, \citenamefont {Norris}, \citenamefont {Blais}, \citenamefont {Krinner},\ and\ \citenamefont {Wallraff}}]{Swiadek:24}%
  \BibitemOpen
  \bibfield  {author} {\bibinfo {author} {\bibfnamefont {F.~m.~c.}\ \bibnamefont {Swiadek}}, \bibinfo {author} {\bibfnamefont {R.}~\bibnamefont {Shillito}}, \bibinfo {author} {\bibfnamefont {P.}~\bibnamefont {Magnard}}, \bibinfo {author} {\bibfnamefont {A.}~\bibnamefont {Remm}}, \bibinfo {author} {\bibfnamefont {C.}~\bibnamefont {Hellings}}, \bibinfo {author} {\bibfnamefont {N.}~\bibnamefont {Lacroix}}, \bibinfo {author} {\bibfnamefont {Q.}~\bibnamefont {Ficheux}}, \bibinfo {author} {\bibfnamefont {D.~C.}\ \bibnamefont {Zanuz}}, \bibinfo {author} {\bibfnamefont {G.~J.}\ \bibnamefont {Norris}}, \bibinfo {author} {\bibfnamefont {A.}~\bibnamefont {Blais}}, \bibinfo {author} {\bibfnamefont {S.}~\bibnamefont {Krinner}},\ and\ \bibinfo {author} {\bibfnamefont {A.}~\bibnamefont {Wallraff}},\ }\bibfield  {title} {\bibinfo {title} {Enhancing dispersive readout of superconducting qubits through dynamic control of the dispersive shift: Experiment and theory},\ }\href {https://doi.org/10.1103/PRXQuantum.5.040326} {\bibfield  {journal} {\bibinfo  {journal} {PRX Quantum}\ }\textbf {\bibinfo {volume} {5}},\ \bibinfo {pages} {040326} (\bibinfo {year} {2024})}\BibitemShut {NoStop}%
\bibitem [{\citenamefont {Yen}\ \emph {et~al.}(2025)\citenamefont {Yen}, \citenamefont {Ye}, \citenamefont {Peng}, \citenamefont {Wang}, \citenamefont {Cunningham}, \citenamefont {Gingras}, \citenamefont {Niedzielski}, \citenamefont {Stickler}, \citenamefont {Serniak}, \citenamefont {Schwartz},\ and\ \citenamefont {O'Brien}}]{Yen:25}%
  \BibitemOpen
  \bibfield  {author} {\bibinfo {author} {\bibfnamefont {A.}~\bibnamefont {Yen}}, \bibinfo {author} {\bibfnamefont {Y.}~\bibnamefont {Ye}}, \bibinfo {author} {\bibfnamefont {K.}~\bibnamefont {Peng}}, \bibinfo {author} {\bibfnamefont {J.}~\bibnamefont {Wang}}, \bibinfo {author} {\bibfnamefont {G.}~\bibnamefont {Cunningham}}, \bibinfo {author} {\bibfnamefont {M.}~\bibnamefont {Gingras}}, \bibinfo {author} {\bibfnamefont {B.~M.}\ \bibnamefont {Niedzielski}}, \bibinfo {author} {\bibfnamefont {H.}~\bibnamefont {Stickler}}, \bibinfo {author} {\bibfnamefont {K.}~\bibnamefont {Serniak}}, \bibinfo {author} {\bibfnamefont {M.~E.}\ \bibnamefont {Schwartz}},\ and\ \bibinfo {author} {\bibfnamefont {K.~P.}\ \bibnamefont {O'Brien}},\ }\bibfield  {title} {\bibinfo {title} {Interferometric purcell suppression of spontaneous emission in a superconducting qubit},\ }\href {https://doi.org/10.1103/PhysRevApplied.23.024068} {\bibfield  {journal} {\bibinfo  {journal} {Phys. Rev. Appl.}\ }\textbf {\bibinfo {volume} {23}},\ \bibinfo {pages} {024068} (\bibinfo {year} {2025})}\BibitemShut {NoStop}%
\bibitem [{\citenamefont {Jeffrey}\ \emph {et~al.}(2014)\citenamefont {Jeffrey}, \citenamefont {Sank}, \citenamefont {Mutus}, \citenamefont {White}, \citenamefont {Kelly}, \citenamefont {Barends}, \citenamefont {Chen}, \citenamefont {Chen}, \citenamefont {Chiaro}, \citenamefont {Dunsworth}, \citenamefont {Megrant}, \citenamefont {O'Malley}, \citenamefont {Neill}, \citenamefont {Roushan}, \citenamefont {Vainsencher}, \citenamefont {Wenner}, \citenamefont {Cleland},\ and\ \citenamefont {Martinis}}]{Jeffrey:14}%
  \BibitemOpen
  \bibfield  {author} {\bibinfo {author} {\bibfnamefont {E.}~\bibnamefont {Jeffrey}}, \bibinfo {author} {\bibfnamefont {D.}~\bibnamefont {Sank}}, \bibinfo {author} {\bibfnamefont {J.~Y.}\ \bibnamefont {Mutus}}, \bibinfo {author} {\bibfnamefont {T.~C.}\ \bibnamefont {White}}, \bibinfo {author} {\bibfnamefont {J.}~\bibnamefont {Kelly}}, \bibinfo {author} {\bibfnamefont {R.}~\bibnamefont {Barends}}, \bibinfo {author} {\bibfnamefont {Y.}~\bibnamefont {Chen}}, \bibinfo {author} {\bibfnamefont {Z.}~\bibnamefont {Chen}}, \bibinfo {author} {\bibfnamefont {B.}~\bibnamefont {Chiaro}}, \bibinfo {author} {\bibfnamefont {A.}~\bibnamefont {Dunsworth}}, \bibinfo {author} {\bibfnamefont {A.}~\bibnamefont {Megrant}}, \bibinfo {author} {\bibfnamefont {P.~J.~J.}\ \bibnamefont {O'Malley}}, \bibinfo {author} {\bibfnamefont {C.}~\bibnamefont {Neill}}, \bibinfo {author} {\bibfnamefont {P.}~\bibnamefont {Roushan}}, \bibinfo {author} {\bibfnamefont {A.}~\bibnamefont {Vainsencher}}, \bibinfo {author} {\bibfnamefont {J.}~\bibnamefont {Wenner}}, \bibinfo {author} {\bibfnamefont {A.~N.}\ \bibnamefont {Cleland}},\ and\ \bibinfo {author} {\bibfnamefont {J.~M.}\ \bibnamefont {Martinis}},\ }\bibfield  {title} {\bibinfo {title} {Fast accurate state measurement with superconducting qubits},\ }\href {https://doi.org/10.1103/PhysRevLett.112.190504} {\bibfield  {journal} {\bibinfo  {journal} {Phys. Rev. Lett.}\ }\textbf {\bibinfo {volume} {112}},\ \bibinfo {pages} {190504} (\bibinfo {year} {2014})}\BibitemShut {NoStop}%
\bibitem [{\citenamefont {Bronn}\ \emph {et~al.}(2015{\natexlab{b}})\citenamefont {Bronn}, \citenamefont {Liu}, \citenamefont {Hertzberg}, \citenamefont {Córcoles}, \citenamefont {Houck}, \citenamefont {Gambetta},\ and\ \citenamefont {Chow}}]{Bronn:15-2}%
  \BibitemOpen
  \bibfield  {author} {\bibinfo {author} {\bibfnamefont {N.~T.}\ \bibnamefont {Bronn}}, \bibinfo {author} {\bibfnamefont {Y.}~\bibnamefont {Liu}}, \bibinfo {author} {\bibfnamefont {J.~B.}\ \bibnamefont {Hertzberg}}, \bibinfo {author} {\bibfnamefont {A.~D.}\ \bibnamefont {Córcoles}}, \bibinfo {author} {\bibfnamefont {A.~A.}\ \bibnamefont {Houck}}, \bibinfo {author} {\bibfnamefont {J.~M.}\ \bibnamefont {Gambetta}},\ and\ \bibinfo {author} {\bibfnamefont {J.~M.}\ \bibnamefont {Chow}},\ }\bibfield  {title} {\bibinfo {title} {Broadband filters for abatement of spontaneous emission in circuit quantum electrodynamics},\ }\href {https://doi.org/10.1063/1.4934867} {\bibfield  {journal} {\bibinfo  {journal} {Applied Physics Letters}\ }\textbf {\bibinfo {volume} {107}},\ \bibinfo {pages} {172601} (\bibinfo {year} {2015}{\natexlab{b}})}\BibitemShut {NoStop}%
\bibitem [{\citenamefont {Bronn}\ \emph {et~al.}(2017)\citenamefont {Bronn}, \citenamefont {Abdo}, \citenamefont {Inoue}, \citenamefont {Lekuch}, \citenamefont {C\'orcoles}, \citenamefont {Hertzberg}, \citenamefont {Takita}, \citenamefont {Bishop}, \citenamefont {Gambetta},\ and\ \citenamefont {Chow}}]{Bronn_2017}%
  \BibitemOpen
  \bibfield  {author} {\bibinfo {author} {\bibfnamefont {N.~T.}\ \bibnamefont {Bronn}}, \bibinfo {author} {\bibfnamefont {B.}~\bibnamefont {Abdo}}, \bibinfo {author} {\bibfnamefont {K.}~\bibnamefont {Inoue}}, \bibinfo {author} {\bibfnamefont {S.}~\bibnamefont {Lekuch}}, \bibinfo {author} {\bibfnamefont {A.~D.}\ \bibnamefont {C\'orcoles}}, \bibinfo {author} {\bibfnamefont {J.~B.}\ \bibnamefont {Hertzberg}}, \bibinfo {author} {\bibfnamefont {M.}~\bibnamefont {Takita}}, \bibinfo {author} {\bibfnamefont {L.~S.}\ \bibnamefont {Bishop}}, \bibinfo {author} {\bibfnamefont {J.~M.}\ \bibnamefont {Gambetta}},\ and\ \bibinfo {author} {\bibfnamefont {J.~M.}\ \bibnamefont {Chow}},\ }\bibfield  {title} {\bibinfo {title} {Fast, high-fidelity readout of multiple qubits},\ }\href {https://doi.org/10.1088/1742-6596/834/1/012003} {\bibfield  {journal} {\bibinfo  {journal} {Journal of Physics: Conference Series}\ }\textbf {\bibinfo {volume} {834}},\ \bibinfo {pages} {012003} (\bibinfo {year} {2017})}\BibitemShut {NoStop}%
\bibitem [{\citenamefont {Yan}\ \emph {et~al.}(2023)\citenamefont {Yan}, \citenamefont {Wu}, \citenamefont {Lingenfelter}, \citenamefont {Joshi}, \citenamefont {Andersson}, \citenamefont {Conner}, \citenamefont {Chou}, \citenamefont {Grebel}, \citenamefont {Miller}, \citenamefont {Povey}, \citenamefont {Qiao}, \citenamefont {Clerk},\ and\ \citenamefont {Cleland}}]{Yan:23}%
  \BibitemOpen
  \bibfield  {author} {\bibinfo {author} {\bibfnamefont {H.}~\bibnamefont {Yan}}, \bibinfo {author} {\bibfnamefont {X.}~\bibnamefont {Wu}}, \bibinfo {author} {\bibfnamefont {A.}~\bibnamefont {Lingenfelter}}, \bibinfo {author} {\bibfnamefont {Y.~J.}\ \bibnamefont {Joshi}}, \bibinfo {author} {\bibfnamefont {G.}~\bibnamefont {Andersson}}, \bibinfo {author} {\bibfnamefont {C.~R.}\ \bibnamefont {Conner}}, \bibinfo {author} {\bibfnamefont {M.-H.}\ \bibnamefont {Chou}}, \bibinfo {author} {\bibfnamefont {J.}~\bibnamefont {Grebel}}, \bibinfo {author} {\bibfnamefont {J.~M.}\ \bibnamefont {Miller}}, \bibinfo {author} {\bibfnamefont {R.~G.}\ \bibnamefont {Povey}}, \bibinfo {author} {\bibfnamefont {H.}~\bibnamefont {Qiao}}, \bibinfo {author} {\bibfnamefont {A.~A.}\ \bibnamefont {Clerk}},\ and\ \bibinfo {author} {\bibfnamefont {A.~N.}\ \bibnamefont {Cleland}},\ }\bibfield  {title} {\bibinfo {title} {Broadband bandpass purcell filter for circuit quantum electrodynamics},\ }\href {https://doi.org/10.1063/5.0161893} {\bibfield  {journal} {\bibinfo  {journal} {Applied Physics Letters}\ }\textbf {\bibinfo {volume} {123}},\ \bibinfo {pages} {134001} (\bibinfo {year} {2023})}\BibitemShut {NoStop}%
\bibitem [{\citenamefont {Zhou}\ \emph {et~al.}(2024)\citenamefont {Zhou}, \citenamefont {Cai}, \citenamefont {Zheng}, \citenamefont {Zhou}, \citenamefont {Wang}, \citenamefont {Xiong},\ and\ \citenamefont {Feng}}]{Zhou:24}%
  \BibitemOpen
  \bibfield  {author} {\bibinfo {author} {\bibfnamefont {Y.}~\bibnamefont {Zhou}}, \bibinfo {author} {\bibfnamefont {X.}~\bibnamefont {Cai}}, \bibinfo {author} {\bibfnamefont {Y.}~\bibnamefont {Zheng}}, \bibinfo {author} {\bibfnamefont {B.}~\bibnamefont {Zhou}}, \bibinfo {author} {\bibfnamefont {Y.}~\bibnamefont {Wang}}, \bibinfo {author} {\bibfnamefont {K.}~\bibnamefont {Xiong}},\ and\ \bibinfo {author} {\bibfnamefont {J.}~\bibnamefont {Feng}},\ }\bibfield  {title} {\bibinfo {title} {High-suppression-ratio and wide bandwidth four-stage purcell filter for multiplexed superconducting qubit readout},\ }\href {https://doi.org/10.1063/5.0173539} {\bibfield  {journal} {\bibinfo  {journal} {Journal of Applied Physics}\ }\textbf {\bibinfo {volume} {135}},\ \bibinfo {pages} {024402} (\bibinfo {year} {2024})}\BibitemShut {NoStop}%
\bibitem [{\citenamefont {Gu}\ \emph {et~al.}(2026)\citenamefont {Gu}, \citenamefont {Feng}, \citenamefont {Peng}, \citenamefont {Liang}, \citenamefont {He}, \citenamefont {Xiao}, \citenamefont {Wang}, \citenamefont {Yan}, \citenamefont {Chen}, \citenamefont {Mei}, \citenamefont {Bu}, \citenamefont {Zhang}, \citenamefont {Song}, \citenamefont {Deng}, \citenamefont {Shi}, \citenamefont {Song}, \citenamefont {Zheng}, \citenamefont {Xu}, \citenamefont {Xiang},\ and\ \citenamefont {Fan}}]{Xu:25}%
  \BibitemOpen
  \bibfield  {author} {\bibinfo {author} {\bibfnamefont {X.-Y.}\ \bibnamefont {Gu}}, \bibinfo {author} {\bibfnamefont {D.}~\bibnamefont {Feng}}, \bibinfo {author} {\bibfnamefont {Z.-Y.}\ \bibnamefont {Peng}}, \bibinfo {author} {\bibfnamefont {G.-H.}\ \bibnamefont {Liang}}, \bibinfo {author} {\bibfnamefont {Y.}~\bibnamefont {He}}, \bibinfo {author} {\bibfnamefont {Y.}~\bibnamefont {Xiao}}, \bibinfo {author} {\bibfnamefont {M.-C.}\ \bibnamefont {Wang}}, \bibinfo {author} {\bibfnamefont {Y.}~\bibnamefont {Yan}}, \bibinfo {author} {\bibfnamefont {B.-J.}\ \bibnamefont {Chen}}, \bibinfo {author} {\bibfnamefont {Z.-Y.}\ \bibnamefont {Mei}}, \bibinfo {author} {\bibfnamefont {Y.-Z.}\ \bibnamefont {Bu}}, \bibinfo {author} {\bibfnamefont {J.-C.}\ \bibnamefont {Zhang}}, \bibinfo {author} {\bibfnamefont {J.-C.}\ \bibnamefont {Song}}, \bibinfo {author} {\bibfnamefont {C.-L.}\ \bibnamefont {Deng}}, \bibinfo {author} {\bibfnamefont {Y.-H.}\ \bibnamefont {Shi}}, \bibinfo {author} {\bibfnamefont {X.}~\bibnamefont {Song}}, \bibinfo {author} {\bibfnamefont {D.}~\bibnamefont {Zheng}}, \bibinfo {author} {\bibfnamefont {K.}~\bibnamefont {Xu}}, \bibinfo {author} {\bibfnamefont {Z.}~\bibnamefont {Xiang}},\ and\ \bibinfo {author} {\bibfnamefont {H.}~\bibnamefont {Fan}},\ }\bibfield  {title} {\bibinfo {title} {Multimode purcell filter for superconducting-qubit reset and readout with intrinsic purcell protection},\ }\href {https://doi.org/10.1103/k398-k98j} {\bibfield  {journal} {\bibinfo  {journal} {Phys. Rev. Appl.}\ }\textbf {\bibinfo {volume} {25}},\ \bibinfo {pages} {044003} (\bibinfo {year} {2026})}\BibitemShut {NoStop}%
\bibitem [{\citenamefont {Mirhosseini}\ \emph {et~al.}(2018)\citenamefont {Mirhosseini}, \citenamefont {Kim}, \citenamefont {Ferreira}, \citenamefont {Kalaee}, \citenamefont {Sipahigil}, \citenamefont {Keller},\ and\ \citenamefont {Painter}}]{Mirhosseini:18}%
  \BibitemOpen
  \bibfield  {author} {\bibinfo {author} {\bibfnamefont {M.}~\bibnamefont {Mirhosseini}}, \bibinfo {author} {\bibfnamefont {E.}~\bibnamefont {Kim}}, \bibinfo {author} {\bibfnamefont {V.~S.}\ \bibnamefont {Ferreira}}, \bibinfo {author} {\bibfnamefont {M.}~\bibnamefont {Kalaee}}, \bibinfo {author} {\bibfnamefont {A.}~\bibnamefont {Sipahigil}}, \bibinfo {author} {\bibfnamefont {A.~J.}\ \bibnamefont {Keller}},\ and\ \bibinfo {author} {\bibfnamefont {O.}~\bibnamefont {Painter}},\ }\bibfield  {title} {\bibinfo {title} {Superconducting metamaterials for waveguide quantum electrodynamics},\ }\href {https://doi.org/10.1038/s41467-018-06142-z} {\bibfield  {journal} {\bibinfo  {journal} {Nature Communications}\ }\textbf {\bibinfo {volume} {9}},\ \bibinfo {pages} {3706} (\bibinfo {year} {2018})}\BibitemShut {NoStop}%
\bibitem [{\citenamefont {Koch}\ \emph {et~al.}(2007)\citenamefont {Koch}, \citenamefont {Yu}, \citenamefont {Gambetta}, \citenamefont {Houck}, \citenamefont {Schuster}, \citenamefont {Majer}, \citenamefont {Blais}, \citenamefont {Devoret}, \citenamefont {Girvin},\ and\ \citenamefont {Schoelkopf}}]{Koch:07}%
  \BibitemOpen
  \bibfield  {author} {\bibinfo {author} {\bibfnamefont {J.}~\bibnamefont {Koch}}, \bibinfo {author} {\bibfnamefont {T.~M.}\ \bibnamefont {Yu}}, \bibinfo {author} {\bibfnamefont {J.}~\bibnamefont {Gambetta}}, \bibinfo {author} {\bibfnamefont {A.~A.}\ \bibnamefont {Houck}}, \bibinfo {author} {\bibfnamefont {D.~I.}\ \bibnamefont {Schuster}}, \bibinfo {author} {\bibfnamefont {J.}~\bibnamefont {Majer}}, \bibinfo {author} {\bibfnamefont {A.}~\bibnamefont {Blais}}, \bibinfo {author} {\bibfnamefont {M.~H.}\ \bibnamefont {Devoret}}, \bibinfo {author} {\bibfnamefont {S.~M.}\ \bibnamefont {Girvin}},\ and\ \bibinfo {author} {\bibfnamefont {R.~J.}\ \bibnamefont {Schoelkopf}},\ }\bibfield  {title} {\bibinfo {title} {Charge-insensitive qubit design derived from the cooper pair box},\ }\href {https://doi.org/10.1103/PhysRevA.76.042319} {\bibfield  {journal} {\bibinfo  {journal} {Phys. Rev. A}\ }\textbf {\bibinfo {volume} {76}},\ \bibinfo {pages} {042319} (\bibinfo {year} {2007})}\BibitemShut {NoStop}%
\bibitem [{\citenamefont {Orlando}\ \emph {et~al.}(1999)\citenamefont {Orlando}, \citenamefont {Mooij}, \citenamefont {Tian}, \citenamefont {van~der Wal}, \citenamefont {Levitov}, \citenamefont {Lloyd},\ and\ \citenamefont {Mazo}}]{Orlando:99}%
  \BibitemOpen
  \bibfield  {author} {\bibinfo {author} {\bibfnamefont {T.~P.}\ \bibnamefont {Orlando}}, \bibinfo {author} {\bibfnamefont {J.~E.}\ \bibnamefont {Mooij}}, \bibinfo {author} {\bibfnamefont {L.}~\bibnamefont {Tian}}, \bibinfo {author} {\bibfnamefont {C.~H.}\ \bibnamefont {van~der Wal}}, \bibinfo {author} {\bibfnamefont {L.~S.}\ \bibnamefont {Levitov}}, \bibinfo {author} {\bibfnamefont {S.}~\bibnamefont {Lloyd}},\ and\ \bibinfo {author} {\bibfnamefont {J.~J.}\ \bibnamefont {Mazo}},\ }\bibfield  {title} {\bibinfo {title} {Superconducting persistent-current qubit},\ }\href {https://doi.org/10.1103/PhysRevB.60.15398} {\bibfield  {journal} {\bibinfo  {journal} {Phys. Rev. B}\ }\textbf {\bibinfo {volume} {60}},\ \bibinfo {pages} {15398} (\bibinfo {year} {1999})}\BibitemShut {NoStop}%
\bibitem [{\citenamefont {Manucharyan}\ \emph {et~al.}(2009)\citenamefont {Manucharyan}, \citenamefont {Koch}, \citenamefont {Glazman},\ and\ \citenamefont {Devoret}}]{Manucharyan:09}%
  \BibitemOpen
  \bibfield  {author} {\bibinfo {author} {\bibfnamefont {V.~E.}\ \bibnamefont {Manucharyan}}, \bibinfo {author} {\bibfnamefont {J.}~\bibnamefont {Koch}}, \bibinfo {author} {\bibfnamefont {L.~I.}\ \bibnamefont {Glazman}},\ and\ \bibinfo {author} {\bibfnamefont {M.~H.}\ \bibnamefont {Devoret}},\ }\bibfield  {title} {\bibinfo {title} {Fluxonium: Single cooper-pair circuit free of charge offsets},\ }\href {https://doi.org/10.1126/science.1175552} {\bibfield  {journal} {\bibinfo  {journal} {Science}\ }\textbf {\bibinfo {volume} {326}},\ \bibinfo {pages} {113} (\bibinfo {year} {2009})}\BibitemShut {NoStop}%
\bibitem [{\citenamefont {Nguyen}\ \emph {et~al.}(2019)\citenamefont {Nguyen}, \citenamefont {Lin}, \citenamefont {Somoroff}, \citenamefont {Mencia}, \citenamefont {Grabon},\ and\ \citenamefont {Manucharyan}}]{Nguyen:19}%
  \BibitemOpen
  \bibfield  {author} {\bibinfo {author} {\bibfnamefont {L.~B.}\ \bibnamefont {Nguyen}}, \bibinfo {author} {\bibfnamefont {Y.-H.}\ \bibnamefont {Lin}}, \bibinfo {author} {\bibfnamefont {A.}~\bibnamefont {Somoroff}}, \bibinfo {author} {\bibfnamefont {R.}~\bibnamefont {Mencia}}, \bibinfo {author} {\bibfnamefont {N.}~\bibnamefont {Grabon}},\ and\ \bibinfo {author} {\bibfnamefont {V.~E.}\ \bibnamefont {Manucharyan}},\ }\bibfield  {title} {\bibinfo {title} {High-coherence fluxonium qubit},\ }\href {https://doi.org/10.1103/PhysRevX.9.041041} {\bibfield  {journal} {\bibinfo  {journal} {Phys. Rev. X}\ }\textbf {\bibinfo {volume} {9}},\ \bibinfo {pages} {041041} (\bibinfo {year} {2019})}\BibitemShut {NoStop}%
\bibitem [{\citenamefont {Ding}\ \emph {et~al.}(2023)\citenamefont {Ding}, \citenamefont {Hays}, \citenamefont {Sung}, \citenamefont {Kannan}, \citenamefont {An}, \citenamefont {Di~Paolo}, \citenamefont {Karamlou}, \citenamefont {Hazard}, \citenamefont {Azar}, \citenamefont {Kim}, \citenamefont {Niedzielski}, \citenamefont {Melville}, \citenamefont {Schwartz}, \citenamefont {Yoder}, \citenamefont {Orlando}, \citenamefont {Gustavsson}, \citenamefont {Grover}, \citenamefont {Serniak},\ and\ \citenamefont {Oliver}}]{Ding:23}%
  \BibitemOpen
  \bibfield  {author} {\bibinfo {author} {\bibfnamefont {L.}~\bibnamefont {Ding}}, \bibinfo {author} {\bibfnamefont {M.}~\bibnamefont {Hays}}, \bibinfo {author} {\bibfnamefont {Y.}~\bibnamefont {Sung}}, \bibinfo {author} {\bibfnamefont {B.}~\bibnamefont {Kannan}}, \bibinfo {author} {\bibfnamefont {J.}~\bibnamefont {An}}, \bibinfo {author} {\bibfnamefont {A.}~\bibnamefont {Di~Paolo}}, \bibinfo {author} {\bibfnamefont {A.~H.}\ \bibnamefont {Karamlou}}, \bibinfo {author} {\bibfnamefont {T.~M.}\ \bibnamefont {Hazard}}, \bibinfo {author} {\bibfnamefont {K.}~\bibnamefont {Azar}}, \bibinfo {author} {\bibfnamefont {D.~K.}\ \bibnamefont {Kim}}, \bibinfo {author} {\bibfnamefont {B.~M.}\ \bibnamefont {Niedzielski}}, \bibinfo {author} {\bibfnamefont {A.}~\bibnamefont {Melville}}, \bibinfo {author} {\bibfnamefont {M.~E.}\ \bibnamefont {Schwartz}}, \bibinfo {author} {\bibfnamefont {J.~L.}\ \bibnamefont {Yoder}}, \bibinfo {author} {\bibfnamefont {T.~P.}\ \bibnamefont {Orlando}}, \bibinfo {author} {\bibfnamefont {S.}~\bibnamefont {Gustavsson}}, \bibinfo {author} {\bibfnamefont {J.~A.}\ \bibnamefont {Grover}}, \bibinfo {author} {\bibfnamefont {K.}~\bibnamefont {Serniak}},\ and\ \bibinfo {author} {\bibfnamefont {W.~D.}\ \bibnamefont {Oliver}},\ }\bibfield  {title} {\bibinfo {title} {High-fidelity, frequency-flexible two-qubit fluxonium gates with a transmon coupler},\ }\href {https://doi.org/10.1103/PhysRevX.13.031035} {\bibfield  {journal} {\bibinfo  {journal} {Phys. Rev. X}\ }\textbf {\bibinfo {volume} {13}},\ \bibinfo {pages} {031035} (\bibinfo {year} {2023})}\BibitemShut {NoStop}%
\bibitem [{\citenamefont {Rower}\ \emph {et~al.}(2024)\citenamefont {Rower}, \citenamefont {Ding}, \citenamefont {Zhang}, \citenamefont {Hays}, \citenamefont {An}, \citenamefont {Harrington}, \citenamefont {Rosen}, \citenamefont {Gertler}, \citenamefont {Hazard}, \citenamefont {Niedzielski}, \citenamefont {Schwartz}, \citenamefont {Gustavsson}, \citenamefont {Serniak}, \citenamefont {Grover},\ and\ \citenamefont {Oliver}}]{Rower:24}%
  \BibitemOpen
  \bibfield  {author} {\bibinfo {author} {\bibfnamefont {D.~A.}\ \bibnamefont {Rower}}, \bibinfo {author} {\bibfnamefont {L.}~\bibnamefont {Ding}}, \bibinfo {author} {\bibfnamefont {H.}~\bibnamefont {Zhang}}, \bibinfo {author} {\bibfnamefont {M.}~\bibnamefont {Hays}}, \bibinfo {author} {\bibfnamefont {J.}~\bibnamefont {An}}, \bibinfo {author} {\bibfnamefont {P.~M.}\ \bibnamefont {Harrington}}, \bibinfo {author} {\bibfnamefont {I.~T.}\ \bibnamefont {Rosen}}, \bibinfo {author} {\bibfnamefont {J.~M.}\ \bibnamefont {Gertler}}, \bibinfo {author} {\bibfnamefont {T.~M.}\ \bibnamefont {Hazard}}, \bibinfo {author} {\bibfnamefont {B.~M.}\ \bibnamefont {Niedzielski}}, \bibinfo {author} {\bibfnamefont {M.~E.}\ \bibnamefont {Schwartz}}, \bibinfo {author} {\bibfnamefont {S.}~\bibnamefont {Gustavsson}}, \bibinfo {author} {\bibfnamefont {K.}~\bibnamefont {Serniak}}, \bibinfo {author} {\bibfnamefont {J.~A.}\ \bibnamefont {Grover}},\ and\ \bibinfo {author} {\bibfnamefont {W.~D.}\ \bibnamefont {Oliver}},\ }\bibfield  {title} {\bibinfo {title} {Suppressing counter-rotating errors for fast single-qubit gates with fluxonium},\ }\href {https://doi.org/10.1103/PRXQuantum.5.040342} {\bibfield  {journal} {\bibinfo  {journal} {PRX Quantum}\ }\textbf {\bibinfo {volume} {5}},\ \bibinfo {pages} {040342} (\bibinfo {year} {2024})}\BibitemShut {NoStop}%
\bibitem [{\citenamefont {Teoh}\ \emph {et~al.}(2023)\citenamefont {Teoh}, \citenamefont {Winkel}, \citenamefont {Babla}, \citenamefont {Chapman}, \citenamefont {Claes}, \citenamefont {de~Graaf}, \citenamefont {Garmon}, \citenamefont {Kalfus}, \citenamefont {Lu}, \citenamefont {Maiti}, \citenamefont {Sahay}, \citenamefont {Thakur}, \citenamefont {Tsunoda}, \citenamefont {Xue}, \citenamefont {Frunzio}, \citenamefont {Girvin}, \citenamefont {Puri},\ and\ \citenamefont {Schoelkopf}}]{Teoh:23}%
  \BibitemOpen
  \bibfield  {author} {\bibinfo {author} {\bibfnamefont {J.~D.}\ \bibnamefont {Teoh}}, \bibinfo {author} {\bibfnamefont {P.}~\bibnamefont {Winkel}}, \bibinfo {author} {\bibfnamefont {H.~K.}\ \bibnamefont {Babla}}, \bibinfo {author} {\bibfnamefont {B.~J.}\ \bibnamefont {Chapman}}, \bibinfo {author} {\bibfnamefont {J.}~\bibnamefont {Claes}}, \bibinfo {author} {\bibfnamefont {S.~J.}\ \bibnamefont {de~Graaf}}, \bibinfo {author} {\bibfnamefont {J.~W.~O.}\ \bibnamefont {Garmon}}, \bibinfo {author} {\bibfnamefont {W.~D.}\ \bibnamefont {Kalfus}}, \bibinfo {author} {\bibfnamefont {Y.}~\bibnamefont {Lu}}, \bibinfo {author} {\bibfnamefont {A.}~\bibnamefont {Maiti}}, \bibinfo {author} {\bibfnamefont {K.}~\bibnamefont {Sahay}}, \bibinfo {author} {\bibfnamefont {N.}~\bibnamefont {Thakur}}, \bibinfo {author} {\bibfnamefont {T.}~\bibnamefont {Tsunoda}}, \bibinfo {author} {\bibfnamefont {S.~H.}\ \bibnamefont {Xue}}, \bibinfo {author} {\bibfnamefont {L.}~\bibnamefont {Frunzio}}, \bibinfo {author} {\bibfnamefont {S.~M.}\ \bibnamefont {Girvin}}, \bibinfo {author} {\bibfnamefont {S.}~\bibnamefont {Puri}},\ and\ \bibinfo {author} {\bibfnamefont {R.~J.}\ \bibnamefont {Schoelkopf}},\ }\bibfield  {title} {\bibinfo {title} {Dual-rail encoding with superconducting cavities},\ }\href {https://doi.org/10.1073/pnas.2221736120} {\bibfield  {journal} {\bibinfo  {journal} {Proceedings of the National Academy of Sciences}\ }\textbf {\bibinfo {volume} {120}},\ \bibinfo {pages} {e2221736120} (\bibinfo {year} {2023})}\BibitemShut {NoStop}%
\bibitem [{\citenamefont {Levine}\ \emph {et~al.}(2024)\citenamefont {Levine}, \citenamefont {Haim}, \citenamefont {Hung}, \citenamefont {Alidoust}, \citenamefont {Kalaee}, \citenamefont {DeLorenzo}, \citenamefont {Wollack}, \citenamefont {Arrangoiz-Arriola}, \citenamefont {Khalajhedayati}, \citenamefont {Sanil}, \citenamefont {Moradinejad}, \citenamefont {Vaknin}, \citenamefont {Kubica}, \citenamefont {Hover}, \citenamefont {Aghaeimeibodi}, \citenamefont {Alcid}, \citenamefont {Baek}, \citenamefont {Barnett}, \citenamefont {Bawdekar}, \citenamefont {Bienias}, \citenamefont {Carson}, \citenamefont {Chen}, \citenamefont {Chen}, \citenamefont {Chinkezian}, \citenamefont {Chisholm}, \citenamefont {Clifford}, \citenamefont {Cosmic}, \citenamefont {Crisosto}, \citenamefont {Dalzell}, \citenamefont {Davis}, \citenamefont {D'Ewart}, \citenamefont {Diez}, \citenamefont {D'Souza}, \citenamefont {Dumitrescu}, \citenamefont {Elkhouly}, \citenamefont {Fang}, \citenamefont {Fang}, \citenamefont {Flammia}, \citenamefont {Fling}, \citenamefont {Garcia}, \citenamefont {Gharzai}, \citenamefont {Gorshkov}, \citenamefont {Gray}, \citenamefont {Grimberg}, \citenamefont {Grimsmo}, \citenamefont {Hann}, \citenamefont {He}, \citenamefont {Heidel}, \citenamefont {Howell}, \citenamefont {Hunt}, \citenamefont {Iverson}, \citenamefont {Jarrige}, \citenamefont {Jiang}, \citenamefont {Jones}, \citenamefont {Karabalin}, \citenamefont {Karalekas}, \citenamefont {Keller}, \citenamefont {Lasi}, \citenamefont {Lee}, \citenamefont {Ly}, \citenamefont {MacCabe}, \citenamefont {Mahuli}, \citenamefont {Marcaud}, \citenamefont {Matheny}, \citenamefont {McArdle}, \citenamefont {McCabe}, \citenamefont {Merton}, \citenamefont {Miles}, \citenamefont {Milsted}, \citenamefont {Mishra}, \citenamefont {Moncelsi}, \citenamefont {Naghiloo}, \citenamefont {Noh}, \citenamefont {Oblepias}, \citenamefont {Ortuno}, \citenamefont {Owens}, \citenamefont {Pagdilao}, \citenamefont {Panduro}, \citenamefont {Paquette}, \citenamefont {Patel}, \citenamefont {Peairs}, \citenamefont {Perello}, \citenamefont {Peterson}, \citenamefont {Ponte}, \citenamefont {Putterman}, \citenamefont {Refael}, \citenamefont {Reinhold}, \citenamefont {Resnick}, \citenamefont {Reyna}, \citenamefont {Rodriguez}, \citenamefont {Rose}, \citenamefont {Rubin}, \citenamefont {Runyan}, \citenamefont {Ryan}, \citenamefont {Sahmoud}, \citenamefont {Scaffidi}, \citenamefont {Shah}, \citenamefont {Siavoshi}, \citenamefont {Sivarajah}, \citenamefont {Skogland}, \citenamefont {Su}, \citenamefont {Swenson}, \citenamefont {Sylvia}, \citenamefont {Teo}, \citenamefont {Tomada}, \citenamefont {Torlai}, \citenamefont {Wistrom}, \citenamefont {Zhang}, \citenamefont {Zuk}, \citenamefont {Clerk}, \citenamefont {Brand\~ao}, \citenamefont {Retzker},\ and\ \citenamefont {Painter}}]{Levine:24}%
  \BibitemOpen
  \bibfield  {author} {\bibinfo {author} {\bibfnamefont {H.}~\bibnamefont {Levine}}, \bibinfo {author} {\bibfnamefont {A.}~\bibnamefont {Haim}}, \bibinfo {author} {\bibfnamefont {J.~S.~C.}\ \bibnamefont {Hung}}, \bibinfo {author} {\bibfnamefont {N.}~\bibnamefont {Alidoust}}, \bibinfo {author} {\bibfnamefont {M.}~\bibnamefont {Kalaee}}, \bibinfo {author} {\bibfnamefont {L.}~\bibnamefont {DeLorenzo}}, \bibinfo {author} {\bibfnamefont {E.~A.}\ \bibnamefont {Wollack}}, \bibinfo {author} {\bibfnamefont {P.}~\bibnamefont {Arrangoiz-Arriola}}, \bibinfo {author} {\bibfnamefont {A.}~\bibnamefont {Khalajhedayati}}, \bibinfo {author} {\bibfnamefont {R.}~\bibnamefont {Sanil}}, \bibinfo {author} {\bibfnamefont {H.}~\bibnamefont {Moradinejad}}, \bibinfo {author} {\bibfnamefont {Y.}~\bibnamefont {Vaknin}}, \bibinfo {author} {\bibfnamefont {A.}~\bibnamefont {Kubica}}, \bibinfo {author} {\bibfnamefont {D.}~\bibnamefont {Hover}}, \bibinfo {author} {\bibfnamefont {S.}~\bibnamefont {Aghaeimeibodi}}, \bibinfo {author} {\bibfnamefont {J.~A.}\ \bibnamefont {Alcid}}, \bibinfo {author} {\bibfnamefont {C.}~\bibnamefont {Baek}}, \bibinfo {author} {\bibfnamefont {J.}~\bibnamefont {Barnett}}, \bibinfo {author} {\bibfnamefont {K.}~\bibnamefont {Bawdekar}}, \bibinfo {author} {\bibfnamefont {P.}~\bibnamefont {Bienias}}, \bibinfo {author} {\bibfnamefont {H.~A.}\ \bibnamefont {Carson}}, \bibinfo {author} {\bibfnamefont {C.}~\bibnamefont {Chen}}, \bibinfo {author} {\bibfnamefont {L.}~\bibnamefont {Chen}}, \bibinfo {author} {\bibfnamefont {H.}~\bibnamefont {Chinkezian}}, \bibinfo {author} {\bibfnamefont {E.~M.}\ \bibnamefont {Chisholm}}, \bibinfo {author} {\bibfnamefont {A.}~\bibnamefont {Clifford}}, \bibinfo {author} {\bibfnamefont {R.}~\bibnamefont {Cosmic}}, \bibinfo {author} {\bibfnamefont {N.}~\bibnamefont {Crisosto}}, \bibinfo {author} {\bibfnamefont {A.~M.}\ \bibnamefont {Dalzell}}, \bibinfo {author} {\bibfnamefont {E.}~\bibnamefont {Davis}}, \bibinfo {author} {\bibfnamefont {J.~M.}\ \bibnamefont {D'Ewart}}, \bibinfo {author} {\bibfnamefont {S.}~\bibnamefont {Diez}}, \bibinfo {author} {\bibfnamefont {N.}~\bibnamefont {D'Souza}}, \bibinfo {author} {\bibfnamefont {P.~T.}\ \bibnamefont {Dumitrescu}}, \bibinfo {author} {\bibfnamefont {E.}~\bibnamefont {Elkhouly}}, \bibinfo {author} {\bibfnamefont {M.~T.}\ \bibnamefont {Fang}}, \bibinfo {author} {\bibfnamefont {Y.}~\bibnamefont {Fang}}, \bibinfo {author} {\bibfnamefont {S.}~\bibnamefont {Flammia}}, \bibinfo {author} {\bibfnamefont {M.~J.}\ \bibnamefont {Fling}}, \bibinfo {author} {\bibfnamefont {G.}~\bibnamefont {Garcia}}, \bibinfo {author} {\bibfnamefont {M.~K.}\ \bibnamefont {Gharzai}}, \bibinfo {author} {\bibfnamefont {A.~V.}\ \bibnamefont {Gorshkov}}, \bibinfo {author} {\bibfnamefont {M.~J.}\ \bibnamefont {Gray}}, \bibinfo {author} {\bibfnamefont {S.}~\bibnamefont {Grimberg}}, \bibinfo {author} {\bibfnamefont {A.~L.}\ \bibnamefont {Grimsmo}}, \bibinfo {author} {\bibfnamefont {C.~T.}\ \bibnamefont {Hann}}, \bibinfo {author} {\bibfnamefont {Y.}~\bibnamefont {He}}, \bibinfo {author} {\bibfnamefont {S.}~\bibnamefont {Heidel}}, \bibinfo {author} {\bibfnamefont {S.}~\bibnamefont {Howell}}, \bibinfo {author} {\bibfnamefont {M.}~\bibnamefont {Hunt}}, \bibinfo {author} {\bibfnamefont {J.}~\bibnamefont {Iverson}}, \bibinfo {author} {\bibfnamefont {I.}~\bibnamefont {Jarrige}}, \bibinfo {author} {\bibfnamefont {L.}~\bibnamefont {Jiang}}, \bibinfo {author} {\bibfnamefont {W.~M.}\ \bibnamefont {Jones}}, \bibinfo {author} {\bibfnamefont {R.}~\bibnamefont {Karabalin}}, \bibinfo {author} {\bibfnamefont {P.~J.}\ \bibnamefont {Karalekas}}, \bibinfo {author} {\bibfnamefont {A.~J.}\ \bibnamefont {Keller}}, \bibinfo {author} {\bibfnamefont {D.}~\bibnamefont {Lasi}}, \bibinfo {author} {\bibfnamefont {M.}~\bibnamefont {Lee}}, \bibinfo {author} {\bibfnamefont {V.}~\bibnamefont {Ly}}, \bibinfo {author} {\bibfnamefont {G.}~\bibnamefont {MacCabe}}, \bibinfo {author} {\bibfnamefont {N.}~\bibnamefont {Mahuli}}, \bibinfo {author} {\bibfnamefont {G.}~\bibnamefont {Marcaud}}, \bibinfo {author} {\bibfnamefont {M.~H.}\ \bibnamefont {Matheny}}, \bibinfo {author} {\bibfnamefont {S.}~\bibnamefont {McArdle}}, \bibinfo {author} {\bibfnamefont {G.}~\bibnamefont {McCabe}}, \bibinfo {author} {\bibfnamefont {G.}~\bibnamefont {Merton}}, \bibinfo {author} {\bibfnamefont {C.}~\bibnamefont {Miles}}, \bibinfo {author} {\bibfnamefont {A.}~\bibnamefont {Milsted}}, \bibinfo {author} {\bibfnamefont {A.}~\bibnamefont {Mishra}}, \bibinfo {author} {\bibfnamefont {L.}~\bibnamefont {Moncelsi}}, \bibinfo {author} {\bibfnamefont {M.}~\bibnamefont {Naghiloo}}, \bibinfo {author} {\bibfnamefont {K.}~\bibnamefont {Noh}}, \bibinfo {author} {\bibfnamefont {E.}~\bibnamefont {Oblepias}}, \bibinfo {author} {\bibfnamefont {G.}~\bibnamefont {Ortuno}}, \bibinfo {author} {\bibfnamefont {J.~C.}\ \bibnamefont {Owens}}, \bibinfo {author} {\bibfnamefont {J.}~\bibnamefont {Pagdilao}}, \bibinfo {author} {\bibfnamefont {A.}~\bibnamefont {Panduro}}, \bibinfo {author} {\bibfnamefont {J.-P.}\ \bibnamefont {Paquette}}, \bibinfo {author} {\bibfnamefont {R.~N.}\ \bibnamefont {Patel}}, \bibinfo {author} {\bibfnamefont {G.}~\bibnamefont {Peairs}}, \bibinfo {author} {\bibfnamefont {D.~J.}\ \bibnamefont {Perello}}, \bibinfo {author} {\bibfnamefont {E.~C.}\ \bibnamefont {Peterson}}, \bibinfo {author} {\bibfnamefont {S.}~\bibnamefont {Ponte}}, \bibinfo {author} {\bibfnamefont {H.}~\bibnamefont {Putterman}}, \bibinfo {author} {\bibfnamefont {G.}~\bibnamefont {Refael}}, \bibinfo {author} {\bibfnamefont {P.}~\bibnamefont {Reinhold}}, \bibinfo {author} {\bibfnamefont {R.}~\bibnamefont {Resnick}}, \bibinfo {author} {\bibfnamefont {O.~A.}\ \bibnamefont {Reyna}}, \bibinfo {author} {\bibfnamefont {R.}~\bibnamefont {Rodriguez}}, \bibinfo {author} {\bibfnamefont {J.}~\bibnamefont {Rose}}, \bibinfo {author} {\bibfnamefont {A.~H.}\ \bibnamefont {Rubin}}, \bibinfo {author} {\bibfnamefont {M.}~\bibnamefont {Runyan}}, \bibinfo {author} {\bibfnamefont {C.~A.}\ \bibnamefont {Ryan}}, \bibinfo {author} {\bibfnamefont {A.}~\bibnamefont {Sahmoud}}, \bibinfo {author} {\bibfnamefont {T.}~\bibnamefont {Scaffidi}}, \bibinfo {author} {\bibfnamefont {B.}~\bibnamefont {Shah}}, \bibinfo {author} {\bibfnamefont {S.}~\bibnamefont {Siavoshi}}, \bibinfo {author} {\bibfnamefont {P.}~\bibnamefont {Sivarajah}}, \bibinfo {author} {\bibfnamefont {T.}~\bibnamefont {Skogland}}, \bibinfo {author} {\bibfnamefont {C.-J.}\ \bibnamefont {Su}}, \bibinfo {author} {\bibfnamefont {L.~J.}\ \bibnamefont {Swenson}}, \bibinfo {author} {\bibfnamefont {J.}~\bibnamefont {Sylvia}}, \bibinfo {author} {\bibfnamefont {S.~M.}\ \bibnamefont {Teo}}, \bibinfo {author} {\bibfnamefont {A.}~\bibnamefont {Tomada}}, \bibinfo {author} {\bibfnamefont {G.}~\bibnamefont {Torlai}}, \bibinfo {author} {\bibfnamefont {M.}~\bibnamefont {Wistrom}}, \bibinfo {author} {\bibfnamefont {K.}~\bibnamefont {Zhang}}, \bibinfo {author} {\bibfnamefont {I.}~\bibnamefont {Zuk}}, \bibinfo {author} {\bibfnamefont {A.~A.}\ \bibnamefont {Clerk}}, \bibinfo {author} {\bibfnamefont {F.~G. S.~L.}\ \bibnamefont {Brand\~ao}}, \bibinfo {author} {\bibfnamefont {A.}~\bibnamefont {Retzker}},\ and\ \bibinfo {author} {\bibfnamefont {O.}~\bibnamefont {Painter}},\ }\bibfield  {title} {\bibinfo {title} {Demonstrating a long-coherence dual-rail erasure qubit using tunable transmons},\ }\href {https://doi.org/10.1103/PhysRevX.14.011051} {\bibfield  {journal} {\bibinfo  {journal} {Phys. Rev. X}\ }\textbf {\bibinfo {volume} {14}},\ \bibinfo {pages} {011051} (\bibinfo {year} {2024})}\BibitemShut {NoStop}%
\bibitem [{\citenamefont {Mirrahimi}\ \emph {et~al.}(2014)\citenamefont {Mirrahimi}, \citenamefont {Leghtas}, \citenamefont {Albert}, \citenamefont {Touzard}, \citenamefont {Schoelkopf}, \citenamefont {Jiang},\ and\ \citenamefont {Devoret}}]{Mirrahimi:14}%
  \BibitemOpen
  \bibfield  {author} {\bibinfo {author} {\bibfnamefont {M.}~\bibnamefont {Mirrahimi}}, \bibinfo {author} {\bibfnamefont {Z.}~\bibnamefont {Leghtas}}, \bibinfo {author} {\bibfnamefont {V.~V.}\ \bibnamefont {Albert}}, \bibinfo {author} {\bibfnamefont {S.}~\bibnamefont {Touzard}}, \bibinfo {author} {\bibfnamefont {R.~J.}\ \bibnamefont {Schoelkopf}}, \bibinfo {author} {\bibfnamefont {L.}~\bibnamefont {Jiang}},\ and\ \bibinfo {author} {\bibfnamefont {M.~H.}\ \bibnamefont {Devoret}},\ }\bibfield  {title} {\bibinfo {title} {Dynamically protected cat-qubits: a new paradigm for universal quantum computation},\ }\href {https://doi.org/10.1088/1367-2630/16/4/045014} {\bibfield  {journal} {\bibinfo  {journal} {New Journal of Physics}\ }\textbf {\bibinfo {volume} {16}},\ \bibinfo {pages} {045014} (\bibinfo {year} {2014})}\BibitemShut {NoStop}%
\bibitem [{\citenamefont {Ding}\ \emph {et~al.}(2025)\citenamefont {Ding}, \citenamefont {Brock}, \citenamefont {Eickbusch}, \citenamefont {Koottandavida}, \citenamefont {Frattini}, \citenamefont {Corti{\~{n}}as}, \citenamefont {Joshi}, \citenamefont {de~Graaf}, \citenamefont {Chapman}, \citenamefont {Ganjam}, \citenamefont {Frunzio}, \citenamefont {Schoelkopf},\ and\ \citenamefont {Devoret}}]{Ding:25}%
  \BibitemOpen
  \bibfield  {author} {\bibinfo {author} {\bibfnamefont {A.~Z.}\ \bibnamefont {Ding}}, \bibinfo {author} {\bibfnamefont {B.~L.}\ \bibnamefont {Brock}}, \bibinfo {author} {\bibfnamefont {A.}~\bibnamefont {Eickbusch}}, \bibinfo {author} {\bibfnamefont {A.}~\bibnamefont {Koottandavida}}, \bibinfo {author} {\bibfnamefont {N.~E.}\ \bibnamefont {Frattini}}, \bibinfo {author} {\bibfnamefont {R.~G.}\ \bibnamefont {Corti{\~{n}}as}}, \bibinfo {author} {\bibfnamefont {V.~R.}\ \bibnamefont {Joshi}}, \bibinfo {author} {\bibfnamefont {S.~J.}\ \bibnamefont {de~Graaf}}, \bibinfo {author} {\bibfnamefont {B.~J.}\ \bibnamefont {Chapman}}, \bibinfo {author} {\bibfnamefont {S.}~\bibnamefont {Ganjam}}, \bibinfo {author} {\bibfnamefont {L.}~\bibnamefont {Frunzio}}, \bibinfo {author} {\bibfnamefont {R.~J.}\ \bibnamefont {Schoelkopf}},\ and\ \bibinfo {author} {\bibfnamefont {M.~H.}\ \bibnamefont {Devoret}},\ }\bibfield  {title} {\bibinfo {title} {Quantum control of an oscillator with a kerr-cat qubit},\ }\href {https://doi.org/10.1038/s41467-025-60352-w} {\bibfield  {journal} {\bibinfo  {journal} {Nature Communications}\ }\textbf {\bibinfo {volume} {16}},\ \bibinfo {pages} {5279} (\bibinfo {year} {2025})}\BibitemShut {NoStop}%
\bibitem [{\citenamefont {Yamamoto}\ \emph {et~al.}(2008)\citenamefont {Yamamoto}, \citenamefont {Inomata}, \citenamefont {Watanabe}, \citenamefont {Matsuba}, \citenamefont {Miyazaki}, \citenamefont {Oliver}, \citenamefont {Nakamura},\ and\ \citenamefont {Tsai}}]{Yamamoto:08}%
  \BibitemOpen
  \bibfield  {author} {\bibinfo {author} {\bibfnamefont {T.}~\bibnamefont {Yamamoto}}, \bibinfo {author} {\bibfnamefont {K.}~\bibnamefont {Inomata}}, \bibinfo {author} {\bibfnamefont {M.}~\bibnamefont {Watanabe}}, \bibinfo {author} {\bibfnamefont {K.}~\bibnamefont {Matsuba}}, \bibinfo {author} {\bibfnamefont {T.}~\bibnamefont {Miyazaki}}, \bibinfo {author} {\bibfnamefont {W.~D.}\ \bibnamefont {Oliver}}, \bibinfo {author} {\bibfnamefont {Y.}~\bibnamefont {Nakamura}},\ and\ \bibinfo {author} {\bibfnamefont {J.~S.}\ \bibnamefont {Tsai}},\ }\bibfield  {title} {\bibinfo {title} {Flux-driven josephson parametric amplifier},\ }\href {https://doi.org/10.1063/1.2964182} {\bibfield  {journal} {\bibinfo  {journal} {Applied Physics Letters}\ }\textbf {\bibinfo {volume} {93}},\ \bibinfo {pages} {042510} (\bibinfo {year} {2008})}\BibitemShut {NoStop}%
\bibitem [{\citenamefont {Clerk}\ \emph {et~al.}(2010)\citenamefont {Clerk}, \citenamefont {Devoret}, \citenamefont {Girvin}, \citenamefont {Marquardt},\ and\ \citenamefont {Schoelkopf}}]{Clerk:10}%
  \BibitemOpen
  \bibfield  {author} {\bibinfo {author} {\bibfnamefont {A.~A.}\ \bibnamefont {Clerk}}, \bibinfo {author} {\bibfnamefont {M.~H.}\ \bibnamefont {Devoret}}, \bibinfo {author} {\bibfnamefont {S.~M.}\ \bibnamefont {Girvin}}, \bibinfo {author} {\bibfnamefont {F.}~\bibnamefont {Marquardt}},\ and\ \bibinfo {author} {\bibfnamefont {R.~J.}\ \bibnamefont {Schoelkopf}},\ }\bibfield  {title} {\bibinfo {title} {Introduction to quantum noise, measurement, and amplification},\ }\href {https://doi.org/10.1103/RevModPhys.82.1155} {\bibfield  {journal} {\bibinfo  {journal} {Rev. Mod. Phys.}\ }\textbf {\bibinfo {volume} {82}},\ \bibinfo {pages} {1155} (\bibinfo {year} {2010})}\BibitemShut {NoStop}%
\bibitem [{\citenamefont {Kurizki}\ \emph {et~al.}(2015)\citenamefont {Kurizki}, \citenamefont {Bertet}, \citenamefont {Kubo}, \citenamefont {Mølmer}, \citenamefont {Petrosyan}, \citenamefont {Rabl},\ and\ \citenamefont {Schmiedmayer}}]{Kurizki:15}%
  \BibitemOpen
  \bibfield  {author} {\bibinfo {author} {\bibfnamefont {G.}~\bibnamefont {Kurizki}}, \bibinfo {author} {\bibfnamefont {P.}~\bibnamefont {Bertet}}, \bibinfo {author} {\bibfnamefont {Y.}~\bibnamefont {Kubo}}, \bibinfo {author} {\bibfnamefont {K.}~\bibnamefont {Mølmer}}, \bibinfo {author} {\bibfnamefont {D.}~\bibnamefont {Petrosyan}}, \bibinfo {author} {\bibfnamefont {P.}~\bibnamefont {Rabl}},\ and\ \bibinfo {author} {\bibfnamefont {J.}~\bibnamefont {Schmiedmayer}},\ }\bibfield  {title} {\bibinfo {title} {Quantum technologies with hybrid systems},\ }\href {https://doi.org/10.1073/pnas.1419326112} {\bibfield  {journal} {\bibinfo  {journal} {Proceedings of the National Academy of Sciences}\ }\textbf {\bibinfo {volume} {112}},\ \bibinfo {pages} {3866} (\bibinfo {year} {2015})}\BibitemShut {NoStop}%
\bibitem [{\citenamefont {Shen}\ and\ \citenamefont {Fan}(2005)}]{Shen:05}%
  \BibitemOpen
  \bibfield  {author} {\bibinfo {author} {\bibfnamefont {J.-T.}\ \bibnamefont {Shen}}\ and\ \bibinfo {author} {\bibfnamefont {S.}~\bibnamefont {Fan}},\ }\bibfield  {title} {\bibinfo {title} {Coherent single photon transport in a one-dimensional waveguide coupled with superconducting quantum bits},\ }\href {https://doi.org/10.1103/PhysRevLett.95.213001} {\bibfield  {journal} {\bibinfo  {journal} {Phys. Rev. Lett.}\ }\textbf {\bibinfo {volume} {95}},\ \bibinfo {pages} {213001} (\bibinfo {year} {2005})}\BibitemShut {NoStop}%
\bibitem [{\citenamefont {Shen}\ and\ \citenamefont {Fan}(2009)}]{Shen:09}%
  \BibitemOpen
  \bibfield  {author} {\bibinfo {author} {\bibfnamefont {J.-T.}\ \bibnamefont {Shen}}\ and\ \bibinfo {author} {\bibfnamefont {S.}~\bibnamefont {Fan}},\ }\bibfield  {title} {\bibinfo {title} {Theory of single-photon transport in a single-mode waveguide. i. coupling to a cavity containing a two-level atom},\ }\href {https://doi.org/10.1103/PhysRevA.79.023837} {\bibfield  {journal} {\bibinfo  {journal} {Phys. Rev. A}\ }\textbf {\bibinfo {volume} {79}},\ \bibinfo {pages} {023837} (\bibinfo {year} {2009})}\BibitemShut {NoStop}%
\bibitem [{\citenamefont {Levenson-Falk}\ and\ \citenamefont {Shanto}(2025)}]{levenson2025review}%
  \BibitemOpen
  \bibfield  {author} {\bibinfo {author} {\bibfnamefont {E.~M.}\ \bibnamefont {Levenson-Falk}}\ and\ \bibinfo {author} {\bibfnamefont {S.~A.}\ \bibnamefont {Shanto}},\ }\bibfield  {title} {\bibinfo {title} {A review of design concerns in superconducting quantum circuits},\ }\href@noop {} {\bibfield  {journal} {\bibinfo  {journal} {Materials for Quantum Technology}\ }\textbf {\bibinfo {volume} {5}},\ \bibinfo {pages} {022003} (\bibinfo {year} {2025})}\BibitemShut {NoStop}%
\bibitem [{\citenamefont {Kosen}\ \emph {et~al.}(2022)\citenamefont {Kosen}, \citenamefont {Li}, \citenamefont {Rommel}, \citenamefont {Shiri}, \citenamefont {Warren}, \citenamefont {Gr{\"o}nberg}, \citenamefont {Salonen}, \citenamefont {Abad}, \citenamefont {Bizn{\'a}rov{\'a}}, \citenamefont {Caputo} \emph {et~al.}}]{kosen2022building}%
  \BibitemOpen
  \bibfield  {author} {\bibinfo {author} {\bibfnamefont {S.}~\bibnamefont {Kosen}}, \bibinfo {author} {\bibfnamefont {H.-X.}\ \bibnamefont {Li}}, \bibinfo {author} {\bibfnamefont {M.}~\bibnamefont {Rommel}}, \bibinfo {author} {\bibfnamefont {D.}~\bibnamefont {Shiri}}, \bibinfo {author} {\bibfnamefont {C.}~\bibnamefont {Warren}}, \bibinfo {author} {\bibfnamefont {L.}~\bibnamefont {Gr{\"o}nberg}}, \bibinfo {author} {\bibfnamefont {J.}~\bibnamefont {Salonen}}, \bibinfo {author} {\bibfnamefont {T.}~\bibnamefont {Abad}}, \bibinfo {author} {\bibfnamefont {J.}~\bibnamefont {Bizn{\'a}rov{\'a}}}, \bibinfo {author} {\bibfnamefont {M.}~\bibnamefont {Caputo}}, \emph {et~al.},\ }\bibfield  {title} {\bibinfo {title} {Building blocks of a flip-chip integrated superconducting quantum processor},\ }\href@noop {} {\bibfield  {journal} {\bibinfo  {journal} {Quantum Science \& Technology}\ }\textbf {\bibinfo {volume} {7}},\ \bibinfo {pages} {035018} (\bibinfo {year} {2022})}\BibitemShut {NoStop}%
\bibitem [{\citenamefont {Savola}(2023)}]{savola2023design}%
  \BibitemOpen
  \bibfield  {author} {\bibinfo {author} {\bibfnamefont {N.}~\bibnamefont {Savola}},\ }\emph {\bibinfo {title} {Design and modelling of long-coherence qubits using energy participation ratios}},\ \href@noop {} {\bibinfo {type} {Master's thesis}},\ \bibinfo  {school} {Aalto University} (\bibinfo {year} {2023})\BibitemShut {NoStop}%
\bibitem [{\citenamefont {Moretti}\ \emph {et~al.}(2025)\citenamefont {Moretti}, \citenamefont {Labranca}, \citenamefont {Campana}, \citenamefont {Carobene}, \citenamefont {Gobbo}, \citenamefont {Castellanos-Beltran}, \citenamefont {Olaya}, \citenamefont {Hopkins}, \citenamefont {Banchi}, \citenamefont {Borghesi} \emph {et~al.}}]{moretti2025transmon}%
  \BibitemOpen
  \bibfield  {author} {\bibinfo {author} {\bibfnamefont {R.}~\bibnamefont {Moretti}}, \bibinfo {author} {\bibfnamefont {D.}~\bibnamefont {Labranca}}, \bibinfo {author} {\bibfnamefont {P.}~\bibnamefont {Campana}}, \bibinfo {author} {\bibfnamefont {R.}~\bibnamefont {Carobene}}, \bibinfo {author} {\bibfnamefont {M.}~\bibnamefont {Gobbo}}, \bibinfo {author} {\bibfnamefont {M.~A.}\ \bibnamefont {Castellanos-Beltran}}, \bibinfo {author} {\bibfnamefont {D.}~\bibnamefont {Olaya}}, \bibinfo {author} {\bibfnamefont {P.~F.}\ \bibnamefont {Hopkins}}, \bibinfo {author} {\bibfnamefont {L.}~\bibnamefont {Banchi}}, \bibinfo {author} {\bibfnamefont {M.}~\bibnamefont {Borghesi}}, \emph {et~al.},\ }\bibfield  {title} {\bibinfo {title} {Transmon qubit modeling and characterization for dark matter search},\ }\href@noop {} {\bibfield  {journal} {\bibinfo  {journal} {IEEE Transactions on Quantum Engineering}\ } (\bibinfo {year} {2025})}\BibitemShut {NoStop}%
\bibitem [{\citenamefont {Pozar}(2012)}]{pozar2012microwave}%
  \BibitemOpen
  \bibfield  {author} {\bibinfo {author} {\bibfnamefont {D.~M.}\ \bibnamefont {Pozar}},\ }\href@noop {} {\emph {\bibinfo {title} {Microwave Engineering}}},\ \bibinfo {edition} {4th}\ ed.\ (\bibinfo  {publisher} {John Wiley \& Sons},\ \bibinfo {address} {Hoboken, NJ},\ \bibinfo {year} {2012})\BibitemShut {NoStop}%
\end{thebibliography}
\end{document}